\documentclass[utf8]{frontiersFPHY} 

\setcitestyle{square} 
\usepackage{url,hyperref,lineno,microtype,subcaption}
\usepackage[onehalfspacing]{setspace}

\def\keyFont{\fontsize{8}{11}\helveticabold }
\def\firstAuthorLast{Aaron Held} 
\def\Authors{Aaron Held\,$^{1,*}$}
\def\Address{$^{1}$Theoretical Physics, Blackett Laboratory, Imperial College, London, SW7 2AZ, U.K.}
\def\corrAuthor{Aaron Held}

\def\corrEmail{\href{mailto:a.held@imperial.ac.uk}{a.held@imperial.ac.uk}}

\usepackage[english]{babel}    
\usepackage{xspace}
\usepackage{microtype}    
\usepackage[utf8]{inputenc}    

\usepackage{amsmath,amssymb,amsfonts,dsfont,bm}     
\allowdisplaybreaks     
\usepackage{mathrsfs}
\usepackage[usenames,dvipsnames]{xcolor}
\usepackage{hyperref}
\usepackage{slashed}
\usepackage{youngtab}
\usepackage{physics}
\usepackage{tensor}
\usepackage{wasysym}

\usepackage{siunitx}
\usepackage{csquotes}

\begin{document}

\onecolumn
\firstpage{1}

\title[Effective asymptotic safety]{
Effective asymptotic safety and its predictive power: Gauge-Yukawa theories} 

\author[\firstAuthorLast ]{\Authors} 
\address{} 
\correspondance{} 

\extraAuth{}

\maketitle

\vspace*{-10pt}
\begin{abstract}
Effective field theory provides a new perspective on the predictive power of Renormalization Group fixed points. Critical trajectories between different fixed points confine the regions of UV-complete, IR-complete, as well as conformal theories. The associated boundary surfaces cannot be crossed by the Renormalization Group flow of any effective field theory. We delineate cases in which the boundary surface acts as an infrared attractor for generic effective field theories.
Gauge-Yukawa theories serve as an example that is both perturbative and of direct phenomenological interest. We identify additional matter fields such that all the observed coupling values of the Standard Model, apart from the Abelian hypercharge, lie within the conformal region. We define a quantitative measure of the predictivity of effective asymptotic safety and demonstrate phenomenological constraints for the associated beyond Standard-Model Yukawa couplings. 

\tiny
 \keyFont{ \section{Keywords:} Renormalization Group, interacting fixed points, effective field theory, (beyond) Standard Model physics, asymptotic safety}
\end{abstract}



\section{Motivation}

Effective field theory (EFT) describes all of high-energy physics remarkably well -- see \cite{Brivio:2017vri} for a review of Standard Model (SM) EFT, and \cite{Donoghue:1994dn} for a well-defined EFT of gravity below the Planck scale. EFTs are solely governed by their field content and symmetries (both global and local). The theory space of all possible realizations of an EFT is spanned by the couplings associated with the (infinite) set of all independent symmetry invariants. A specific realization is characterized by its coupling values at some Renormalization Group (RG) scale. Despite the infinitely many couplings, perturbative and local EFTs are predictive towards the infrared (IR) since the infinite tower of higher-order interactions permitted by symmetries and field content is suppressed by powers of the ratio between experimentally accessible scales and the cutoff scale, i.e., by the RG structure around the free fixed point. A sufficiently high cutoff scale thus gives retrospective insight into the success of perturbatively renormalizable gauge-Yukawa theories such as the SM.

On the other hand, it has been of paramount interest to identify fundamental, i.e., ultraviolet (UV) complete, quantum field theories in which the cutoff can be removed -- first asymptotically free \cite{
Gross:1973id,
Politzer:1973fx,
Cheng:1973nv,
Kalashnikov:1976hr,
Oehme:1984yy,
Kubo:1985up,
Kubo:1988zu,
Zimmermann:2001pq,
Giudice:2014tma, 
Holdom:2014hla,
Pelaggi:2015kna, 
Gies:2015lia, 
Pica:2016krb,
Gies:2016kkk, 
Einhorn:2017jbs,
Hansen:2017pwe,
Badziak:2017wxn,
Gies:2018vwk, 
Gies:2019nij},
and more recently, asymptotically safe \cite{
Litim:2014uca,
Sannino:2014lxa,
Litim:2015iea,
Nielsen:2015una,
Rischke:2015mea,
Intriligator:2015xxa,
Esbensen:2015cjw,
Molgaard:2016bqf,
Bajc:2016efj,
Pelaggi:2017wzr,
Abel:2017ujy,
Christiansen:2017qca,
Bond:2017lnq,
Abel:2017rwl,
Bond:2017suy,
Bond:2017wut,
Bond:2017tbw, 
Bond:2018oco,
Barducci:2018ysr, 
Hiller:2019mou,
Bond:2019npq,
Hiller:2019tvg}
gauge-Yukawa theories. See also \cite{
Mann:2017wzh,
Pelaggi:2017abg,
Antipin:2017ebo,
Bajc:2017xwx,
Sannino:2018suq,
Molinaro:2018kjz,
Cacciapaglia:2018avr,
Abel:2018fls,
Wang:2018yer,
Ryttov:2019aux,
Orlando:2019hte,
Sannino:2019sch,
Cai:2019mtu,
Rey:2019wve,
Bajc:2019ari}
(with potential caveats discussed in \cite{
Holdom:2010qs,
Alanne:2019vuk,
Sannino:2019vuf})
for asymptotically safe gauge-Yukawa theories from resummation at a large number of matter fields and \cite{Cacciapaglia:2020kgq} for a recent review including lattice results. 

The RG flow of asymptotically free theories emanates from a fixed point at which all interactions vanish and the theory exhibits classical scale invariance. 
Asymptotic safety \cite{Weinberg:1980gg} generalizes asymptotic freedom to include UV-complete theories that emanate from (partially) interacting fixed points at which some of the couplings remain finite and the fixed-point theory exhibits quantum not classical scale invariance, cf.~\cite{Wetterich:2019qzx}. 

Quantum scale invariance of asymptotically safe theories (including the special case of classical scale invariance of asymptotically free theories) can entail enhanced predictivity. Close to a fixed point, this predictivity can be quantified by the eigenvalues of the linearized RG flow, i.e., by the number of IR-attractive opposed to IR-repulsive directions in theory space, cf.~e.g.~\cite{Eichhorn:2018yfc} for an introduction. Towards the IR, the RG flow can emanate from the fixed point only along IR-repulsive directions. Hence, the subset of EFTs emanating from the fixed point, referred to as its UV-critical hypersurface, is spanned only by the subset of IR-repulsive directions. On the contrary, IR-attractive directions become predictions of such fundamental theories because their coupling values have to remain fixed to the UV-critical hypersurface. A fundamental theory is predictive whenever the UV-critical surface is finite-dimensional.
All perturbative fixed points -- both free and interacting --  are automatically predictive because perturbative quantum fluctuations are (by definition) too weak to cause classically irrelevant couplings to become IR-repulsive.

The present work is limited to non-gravitational theories. Concerning gravity, a considerable body of evidence, pioneered by \cite{Reuter:1996cp}, suggests the existence of an interacting fixed point for Euclidean quantum gravity, cf.~\cite{Percacci:2017fkn,Reuter:2019byg, Eichhorn:2020mte} for introductory texts. If present, such a fixed point could extend EFTs beyond the Planck scale $\Lambda_\text{Planck}$. Here, we will only be concerned with perturbative EFTs at energies below $\Lambda_\text{Planck}$.
Nevertheless, the Planck scale plays a crucial role.
Most conservatively, it is to be regarded as the unavoidable cutoff scale for any non-gravitational theory. Hence, phenomenological implications of (non-gravitational) asymptotic safety should be discussed in the framework of an EFT that is valid only between $\Lambda_\text{Planck}$ and the scale $\Lambda_\text{NP}$ at which the new physics decouples. Assuming that new physics below the electroweak scale $\Lambda_\text{ew}$ is excluded by collider experiments\footnote{This assumption can be circumvented by very weakly coupled particles, in which case the new-physics scale may lie below the electroweak scale. We will not discuss these cases here.
}, the EFTs of interest are therefore valid over at most 17 orders of magnitude in energy scales, i.e.,
\begin{align}
    10^2\;\text{GeV}\approx\Lambda_\text{ew}\lesssim\Lambda_\text{NP}\lesssim\Lambda_\text{Planck}\approx 10^{19}\,\text{GeV}\;.
\end{align}
This motivates us to explore \emph{effective asymptotic safety}, i.e., the predictivity of RG fixed points over a finite range of scales, cf.~also~\cite{Percacci:2010af,Eichhorn:2017eht,Eichhorn:2018yfc,deAlwis:2019aud}. Moreover, we are interested in the global RG structure encompassing all fixed points available in perturbation theory. Effective -- in comparison to fundamental -- asymptotic safety can alter conclusions about phenomenology as well as about the exclusion of specific models. 
To put the results of this paper in a wider context, we make the following simple observation about the RG flow in the theory space of perturbative gauge-Yukawa theories:
\begin{displayquote}
The respective boundaries of the set of all UV-complete, IR-complete, and both UV- and IR-complete theories constitute hypersurfaces in theory space that cannot be crossed by the RG flow of any EFT. With respect to other directions orthogonal to such a boundary hypersurface, the latter inherits the IR-attractive properties of the fixed points by which it is delimited. In these cases, the entire boundary surface, not just the fixed point, can constitute an IR-attractor and generic EFTs tend to cluster close to it\footnote{We caution that these observations require a truncation of the perturbative series or any other expansion which is sufficiently converged to have revealed all physical fixed points. Otherwise, the statement still applies to the truncated RG flow but might lose its phenomenological significance.}.
\end{displayquote}
Possible proof of this claim in more general settings is beyond the scope of this work and might be provided elsewhere in the future. Besides its potential importance for a structural understanding of the behavior of RG flows, it can have phenomenological implications which, in our opinion, deserve more attention. In the following, we will demonstrate this observation for the case of gauge-Yukawa theories. These make for a particularly suitable example because (i) their fixed-point structure is both rich enough and perturbatively well-controlled \cite{Bond:2016dvk, Bond:2017lnq, Bond:2017suy, Bond:2017tbw, Bond:2018oco, Bond:2019npq} and (ii) they are of direct phenomenological significance as possible extensions of the SM \cite{Bond:2017wut, Barducci:2018ysr, Hiller:2019mou, Hiller:2019tvg}. 

\paragraph*{Synopsis of results:}
\begin{itemize}
\item
In Sec.~\ref{sec:RGstructure}, we review the different phases, i.e., the possible perturbative fixed-point structures, of simple gauge-Yukawa theories identified in \cite{Litim:2014uca,Bond:2016dvk}. This discussion allows us to delineate how the above observation is realized. Readers who are familiar with the fixed-point structure of gauge-Yukawa theories and are not interested in a respective discussion of effective asymptotic safety may want to skip this section.
\item
In Sec.~\ref{sec:noGoForSMsubgroups}, we look at each simple SM subgroup by itself which leads to a transparent understanding of why within perturbation theory: (i) additional matter fields can induce fully IR-attractive interacting fixed points for the non-Abelian SM subgroups while (ii) interacting fixed points with UV-attractive directions are not available and (iii) Abelian subgroups will always remain trivial.
\item
Turning to phenomenological implications, we introduce a novel quantitative measure for the global predictivity of EFTs in Sec.~\ref{sec:perturbativityCriterion}. This effective notion of predictivity applies to (finite-dimensional truncations of) perturbative as well as non-perturbative EFTs, more widely.
\item
In Sec.~\ref{sec:SM}, the SM serves as a first example to demonstrate the predictivity measure. Here, we also conclude that whenever the non-Abelian sectors remain perturbative, the Abelian Landau pole of the SM remains safely beyond the Planck scale.
\item
In Sec.~\ref{sec:newMatter}, we discuss phenomenological conclusions of effective asymptotic safety for extensions of the SM by additional matter fields along the lines of \cite{Litim:2014uca}. We identify specific BSM matter for which all the SM coupling values (apart from the Abelian hypercharge coupling) lie within the conformal region.
\end{itemize}    
We conclude in Sec.~\ref{sec:discussion}. The collection of NLO and NNLO beta-functions required for this work, cf.~\cite{Gross:1973id, Politzer:1973fx, Cheng:1973nv, Caswell:1974gg, Tarasov:1976ef, Jones:1981we, Fischler:1982du, Machacek:1983tz, Machacek:1983fi, Jack:1983sk, Curtright:1979mg, Arason:1991ic, Pickering:2001aq, Luo:2002ti, Luo:2002ey, Mihaila:2012fm, Chetyrkin:2012rz, Bednyakov:2012en, Bednyakov:2013eba, Mihaila:2014caa, Schienbein:2018fsw} for original references, is relegated into Appendices.

\section{RG structure of gauge-Yukawa theories: an EFT point of view}
\label{sec:RGstructure}
Before explicitly discussing the SM and its possible extensions, we briefly review the available fixed-point structures of simple gauge-Yukawa theories previously discussed in \cite{Litim:2014uca, Bond:2016dvk}. This serves as a specific example to characterize the global RG structure, effective asymptotic safety and their significance for generic EFTs. For the purpose of this section, we focus on a simple gauge group for which we denote the squared gauge coupling by $\alpha_g = g^2/(4\pi)^2$, cf.~\cite{Bond:2017lnq} for a generalization to semi-simple gauge groups. 

\begin{figure}
    \centering
    \includegraphics[width=1\columnwidth]{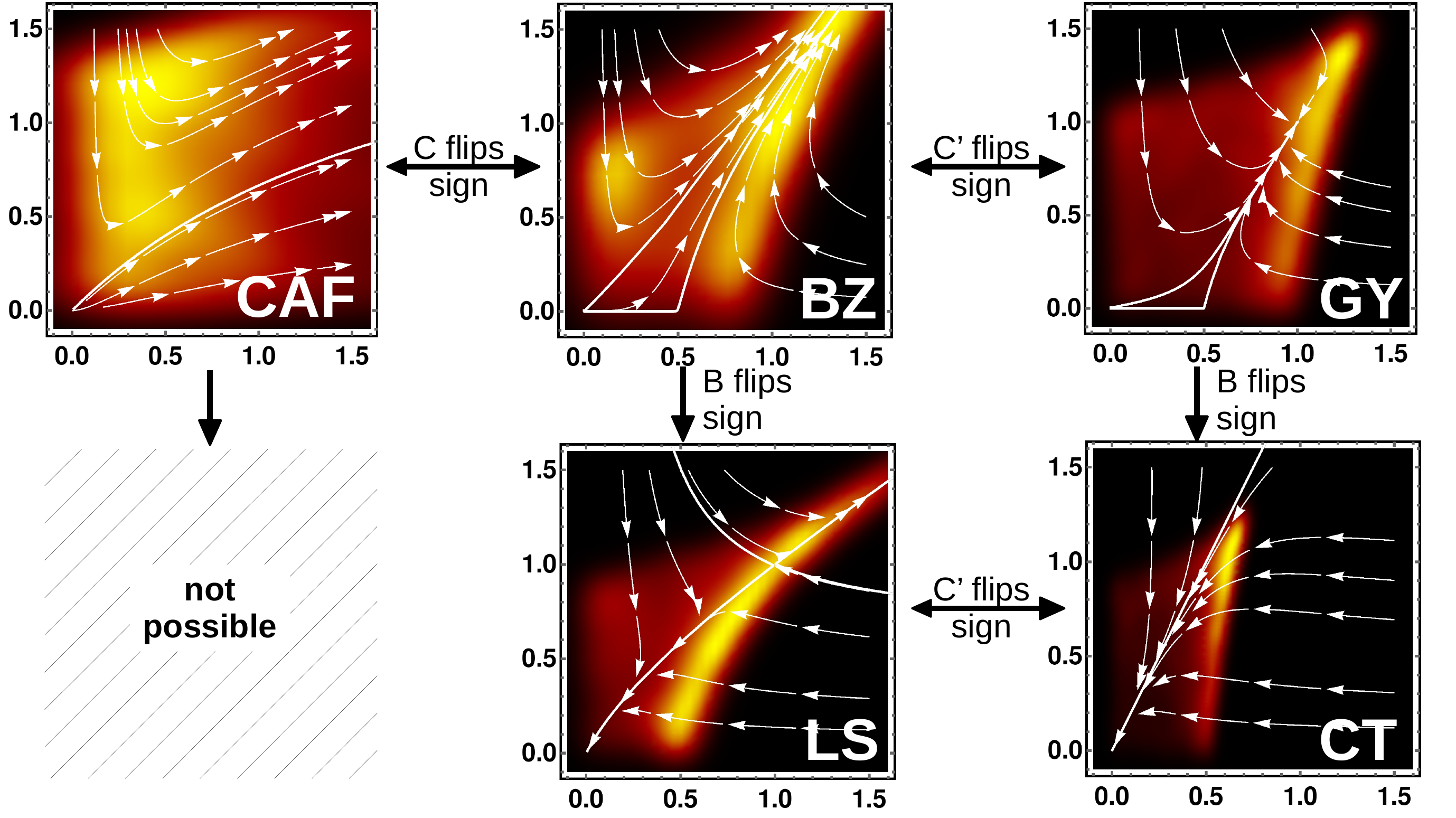}
\caption{
\label{fig:phaseStructure}
Possible RG structures, cf.~\cite{Bond:2016dvk} (see main text for further discussion), of simple gauge Yukawa-theories, depending on the signs of one- and two-loop coefficients $B$, $C$, and $C'$ of the gauge coupling $\beta$-function, cf.~Eq.~\eqref{eq:betaAlphaSimpleGY} and in turn on the gauge group and matter content of the theory. The x-axis (y-axis) shows the gauge coupling $\alpha_g$ (Yukawa coupling $\alpha_y$). Thick white lines indicate the boundary surface of the UV-complete, IR-complete, or conformal regions in theory space. White flow lines (arrows) point towards the IR. The heat maps in the background indicate how a set of EFTs randomly distributed over the full depicted range of couplings is focused towards the boundary surface. Lighter (more yellow) areas indicate a high density of theories below the cutoff scale.
}
\end{figure}

Weyl-consistency conditions suggest that the RG equations of gauge-Yukawa theories should be obtained in hierarchical schemes \cite{Osborn:1989td, Cardy:1988cwa, Jack:1990eb, Osborn:1991gm, Antipin:2013sga, Bond:2017tbw}. In particular, Yukawa couplings contribute to gauge couplings only at 2nd loop order. Quartic couplings contribute to Yukawa and gauge couplings only at 2nd and 3rd loop order, respectively. Therefore, included loop orders of gauge, Yukawa, and quartic couplings should relate as $(n+2,\,n+1,\,n)$, respectively. Throughout this paper, we will neglect quartic couplings for simplicity and work in the $(2,1,0)$-scheme (subsequently referred to as NLO). We check that fixed points remain perturbatively well-controlled by extending to the $(3,2,0)$-scheme (subsequently referred to as NNLO). In the notation of \cite{Bond:2017tbw}, what we call NLO (NNLO) is referred to as NLO$''$ (2NLO$''$). The explicit RG equations of the latter are discussed in Appendices since they merely serve to ensure perturbative control. Following \cite{Bond:2016dvk}, the beta-function of general Yukawa couplings $\mathcal{L}_\text{Yukawa} = -\mathbf{Y}\,\text{tr}\left[\bar{\psi}_L\,\chi\,\psi_R + \bar{\psi}_R\,\chi^\dagger\,\psi_L\right]$, suppressing indices, takes the form
\begin{align}
\label{eq:betaYukawaMatrices}
    \mathbf{\beta}_{\mathbf{Y}} &= \mathbf{E}(\mathbf{Y}) - \alpha\,\mathbf{F}(\mathbf{Y})\;,
\end{align}
where $\mathbf{E}$ and $\mathbf{F}$ are matrices qubic and linear in the Yukawa-coupling matrices $\mathbf{Y}$, respectively. Therefore, besides a trivial fixed point at $\mathbf{Y}_\ast$, additional non-trivial (partial) Yukawa fixed-points exist. The latter depend parametrically on $\alpha_g$ \cite{Bond:2016dvk}, i.e.,
\begin{align}
    \mathbf{Y}_{(\ast)}(\alpha) = \sqrt{\alpha_g}\,\mathbf{C}\;,
    \label{eq:partialYukawaFP}
\end{align}
where the $\mathbf{C}$ is independent of the gauge coupling, cf.~\cite{Bond:2016dvk}. These partial fixed points (also referred to as Yukawa-nullcline) always exist and occur at positive (but not necessarily perturbative) values of the Yukawa couplings. Under the RG flow, they focus the values of Yukawa couplings towards a small IR interval, as for instance in the SM. We will see in Sec.~\ref{sec:newMatter} that they are of phenomenological importance, cf. also \cite{Pendleton:1980as,Hill:1980sq,Wetterich:1981ir}. Evaluating the (2-loop) running of the gauge coupling $\alpha_g$ by use of the above partial fixed-point solution results in
\begin{align}
\label{eq:betaAlphaSimpleGY}
    \beta_{\alpha_g} &= \alpha_g^2\left[
        -B
        - C\,\alpha_g
        + 2D\,\alpha_g
    \right]\;.
\end{align}
The scalar coefficients $B$, $C$, and $D$ are purely group-theoretic and can be found in~\cite{Bond:2016dvk}. $B$ and $C$ arise from gauge-coupling contributions, while $D$ arises from Yukawa couplings at their partial fixed point. Since $\alpha_g = g^2/(4\pi)^2$, the fixed points for $g_\ast$ are physical, i.e., real, only if $\alpha_{g\,\ast}\geqslant 0$. While $D\geqslant0$ ($D=0$ for the vanishing Yukawa fixed point), the signs of $B$ and $C$ depend on the matter content of the theory\footnote{We have chosen the signs to reflect the antiscreening non-Abelian case without matter content. Note that this choice agrees with \cite{Banks:1981nn} but differs from \cite{Litim:2014uca}.}.
Defining $C' = C-2D$ (note that $C'<C$, always), one can fully classify the general theory by two types of interacting fixed points, cf.~\cite{Litim:2014uca, Bond:2016dvk}: one with vanishing and one with non-vanishing Yukawa couplings, i.e.,
\begin{align}
\label{eq:BZFP}
    \alpha_{g\,\ast,\,\text{BZ}}&=-\frac{B}{C}\;,
    \quad\quad\quad
    \mathbf{Y}_{\ast,\,\text{BZ}}=0\;,
    \\
\label{eq:GYFP}
    \alpha_{g\,\ast,\,\text{GY}}&=-\frac{B}{C'},
    \quad\quad\quad
    \mathbf{Y}_{\ast,\,\text{GY}}=\sqrt{\alpha}\,\mathbf{C}\;,
\end{align}
respectively.
Depending on which of these are physical, i.e., occur at $\alpha_{g\,\ast}\geqslant0$, \cite{Bond:2016dvk} have classified the five possible phases, i.e., perturbative fixed-point structures, of simple gauge-Yukawa theories. These are summarized in Table~\ref{tab:phases} and depicted schematically in Figure~\ref{fig:phaseStructure}.

Preceding the EFT discussion of these phases, it is important to distinguish the following terminology. A set of gauge-Yukawa theories is determined by its gauge group and matter content, parameterized, for instance, by the number of fermionic representations $N_F$. To agree with previous literature \cite{Banks:1981nn}, we refer to the possible values of $N_F$ which realize certain gauge-Yukawa phases as `windows'. This is distinct from a particular realization within a set of gauge-Yukawa theories. The latter is further parameterized by coupling values, i.e., by a choice of RG trajectory. When referring to possible values of the couplings, we talk about `regions'. In particular, we say that the set of all UV-complete trajectories makes up the UV-complete region, the set of all IR-complete trajectories makes up the IR-complete region, and the set of all UV- and IR-complete trajectories makes up the `conformal' region. Crucially, the terminology `conformal window' and `conformal region' are to be distinguished.
\begin{table}
\centering
\renewcommand{\arraystretch}{2}
\begin{tabular}{r||c||c|c|c}
       & conditions & $d_\text{IR}$ & $d_\text{UV}$ & $d_\text{UV-IR}$
       \\\hline\hline
       complete asymptotic freedom (\textbf{CAF}) & $B<0$, $C<0$, $C'<0$ & 0 & 2 & 0
       \\\hline
       Banks-Zaks (\textbf{BZ}) conformal window & $B<0$, $C>0$, $C'<0$ & 1 & 2 & 1
       \\\hline
       gauge-Yukawa (\textbf{GY}) conformal window & $B<0$, $C>0$, $C'>0$ & 2 & 2 & 2
       \\\hline
       Litim-Sannino (\textbf{LS}) conformal window & $B>0$, $C>0$, $C'<0$ & 2 & 1 & 1
       \\\hline
       complete triviality (\textbf{CT}) & $B>0$, $C>0$, $C'>0$ & 2 & 0 & 0
\end{tabular}
\caption{
\label{tab:phases}
Different perturbative Renormalization Group phases for simple gauge-Yukawa theories. Conditions on $B$, $C$, and $C'$ can be translated into conditions on parameters like the size of the gauge group and number of fermionic/scalar representations. We also indicate the dimensionalities $d_\text{IR}$, $d_\text{UV}$, and $d_\text{UV-IR} = \text{Min}(d_\text{UV},d_\text{IR})$ of the IR-complete, UV-complete and conformal, i.e., UV- and IR-complete, region in theory space, respectively.
}
\end{table}

\paragraph*{Complete asymptotic freedom (CAF):}
For antiscreening $B>0$ as well as $C>0$, the only physical fixed point is the Gau\ss ian one, cf.~left-hand upper panel in Figure~\ref{fig:phaseStructure}. The former is completely asymptotically free when approached from below the Yukawa nullcline, i.e., whenever $\mathbf{Y}<\mathbf{Y}_\ast(\alpha_g)$. This case also encompasses asymptotic freedom of Yang-Mills theory without Yukawa couplings, cf.~the RG flow along the x-axis in the left-hand upper panel of Figure~\ref{fig:phaseStructure}.
The UV-complete region is 2-dimensional and extends to infinite coupling values (or more accurately beyond perturbative control) although it is partially bound by the Yukawa-nullcline (white line). This boundary surface inherits the IR-attractive property of the free fixed point along the Yukawa direction (y-axis) and is hence IR-attractive from above, i.e., for $\mathbf{Y}>\mathbf{Y}_\ast$. This entails that generic UV-incomplete EFTs will be attracted to the boundary. The IR-complete (and thus also the conformal) region is reduced to the trivial theory. All other theories eventually escape perturbative control towards the IR.

\paragraph*{Banks-Zaks (BZ) conformal window:}
Scalar, as well as fermionic matter, adds screening fluctuations and modifies the running of non-Abelian gauge couplings. This can flip the signs of $C$, $C'$ and $B$. Independent of the specific matter representation, the sign of $C$ is always flipped first and the theory (with vanishing Yukawa couplings) enters the so-called conformal window \cite{Banks:1981nn}, cf.~upper panel in the middle of Figure~\ref{fig:phaseStructure}.
As $C$ flips sign (but before $C'$ or $B$ do so), the Banks-Zaks fixed point becomes physical. 
For vanishing Yukawa coupling and $0<\alpha_g<\alpha_{g\,\ast,\,\text{BZ}}$, the theory is now both UV- and IR-complete. In fact, since the IR-complete and thus the conformal region is still only one-dimensional, the RG-scale can be mapped directly to a unique gauge-coupling value. Put differently, there only exists a single conformal theory.
The gauge-Yukawa fixed point is still not physical and thus every theory with non-vanishing Yukawa coupling will eventually diverge in the IR. This is a consequence of the Banks-Zaks fixed point being IR-attractive in the direction of the gauge coupling but IR-repulsive in the direction of the Yukawa coupling. To distinguish this situation, we refer to this as the `BZ conformal window'.
However, the UV-complete region is still 2-dimensional. Its boundary inherits the partial IR-attractive nature of the two fixed points. (The free fixed point is IR-attractive in the Yukawa-coupling direction and the Banks-Zaks fixed point is IR-attractive along the gauge-coupling direction.) The corresponding sections of the boundary surface act as an IR-attractor, in particular for EFTs outside of the UV-complete region. We emphasize that it is the boundary and not a single fixed point which is IR-attractive.

Adding further matter representations can flip the sign of $C'$ or $B$ first. Therefore, there are now two distinct phases that can occur when further matter is added. Which of these is realized depends on the ratio of scalar and fermionic matter and on the set of possible Yukawa interactions.

\paragraph*{Gauge-Yukawa (GY) conformal window:}
Whenever $C'$ turns negative before $B$ does, the theory develops a fully IR-attractive gauge-Yukawa fixed point, cf.~right-hand upper panel in Figure~\ref{fig:phaseStructure}. As $C'$ is varied, the fixed point formally enters from infinity (or from outside the perturbative regime) along the direction in which the two nullclines of the BZ phase join. The gauge-Yukawa fixed point serves as an endpoint of the two nullclines and delimits the two-dimensional UV-complete region, which is now also IR-complete. As a consequence, there is now a two-dimensional region of distinct conformal theories. In correspondence to the `BZ conformal window,' we refer to this as the `gauge-Yukawa (GY) conformal window'. This case is particularly predictive. If realized only over a finite range of scales, e.g.~due to the decoupling of massive modes, this realizes \emph{effective asymptotic safety}. All EFTs are attracted first to the boundary of the `gauge-Yukawa conformal window' and eventually into the gauge-Yukawa fixed point.

\paragraph*{Litim-Sannino (LS) conformal window:}
If, on the other hand, $B$ turns negative before $C'$ does, a Litim-Sannino fixed point \cite{Litim:2014uca} becomes available, while the Banks-Zaks fixed point \cite{Banks:1981nn} disappears (formally it escapes the perturbative regime in direction of increasing gauge coupling) and the free fixed point becomes fully IR-attractive, cf.~lower panel in the middle of Figure~\ref{fig:phaseStructure}. It is now the IR-complete region which is two-dimensional. However, the UV-complete region and hence the set of conformal theories, is just one-dimensional. The latter is delimited by the free and the Litim-Sannino fixed point, while the former also extends beyond the Litim-Sannino fixed point and corresponds to its UV-critical hypersurface. Concerning generic EFTs, the conformal theory which splits the IR-complete region, i.e., the separatrix between the Litim-Sannino and the free fixed point, acts as an IR-attractor because it inherits this property from the shared IR-attractive direction of both its delimiting fixed points. The boundary of the IR-complete region, however, is \emph{not} IR attractive since it inherits the IR-repulsive direction of the Litim-Sannino fixed point. Again, generic EFTs tend to cluster close to the UV-complete theories, i.e., exhibit \emph{effective asymptotic safety}.

\paragraph*{Complete triviality (CT):}
The final possibility occurs if all three signs are flipped, i.e., $B<0$ and $C<C'<0$. Since all contributions have now turned screening, the theory remains only with the free fixed point. The latter is now fully IR-attractive. This phase occurs for the perturbative range of any Abelian gauge group, cf.~Sec.~\ref{sec:triviality}. Formally, the UV-complete and conformal regions reduce to the trivial theory to which all EFTs are attracted. The IR-complete region now covers all of the theory space. The triviality problem can therefore be seen as a consequence of `effective asymptotic freedom'.

This concludes the review of all possible fixed-point structures \cite{Bond:2016dvk} which can occur due to different cancellations at NLO in simple gauge-Yukawa theories. One can schematically think of semi-simple cases, such as the SM, as the higher-dimensional combinations of these phases, cf.~\cite{Bond:2017lnq} for an explicit discussion.
In the following, we will always check whether potential fixed points persist at NNLO. We present our formal definition of perturbativity in Section~\ref{sec:perturbativityCriterion}. Before doing so, we provide insight into the single gauge groups of the SM which is sufficient to qualitatively understand the available fixed points that we identify in the coupled system in Section~\ref{sec:newMatter}.
In all phases, some form of IR-attractor dominates the RG flow.

\section{Available phases for the simple Standard-Model subgroups}
\label{sec:noGoForSMsubgroups}

Following~\cite{Litim:2014uca}, we remain focused on a simple gauge group with $N_F$ copies of a single type of fermionic representation $R_F$ and uncharged scalars to allow for Yukawa couplings, cf.~App.~\ref{app:generalBetas} or~\cite{Litim:2014uca} for the explicit Lagrangian. In this case, the Yukawa-coupling matrices $\mathbf{E}(\mathbf{Y})$ and $\mathbf{F}(\mathbf{Y})$ in Eq.~\ref{eq:betaYukawaMatrices} reduce to scalar coefficients $E$ and $F$ of a single Yukawa coupling $y$ for which we introduce $\alpha_y = y^2/(4\pi)^2$. The NLO coefficients that determine the interacting fixed-points, cf.~Eqs~\eqref{eq:BZFP}~and~\eqref{eq:GYFP} and the resulting RG-structure are given by, cf.~\cite{Bond:2016dvk}
\begin{align}
    B &= \frac{2}{3\,d^{\text{adj}}} \left(11\,d^{\text{adj}}\,C_2^{\text{adj}}-2\,N_F\, d^{R_F}
   C_2^{R_F}\right)
    \;,\:
    C = \frac{68 \left(C_2^{\text{adj}}\right)^2}{3}-\frac{4 d^{R_F} C_2^{R_F} N_F (5 C_2^{\text{adj}}+3
   C_2^{R_F})}{3 d^{\text{adj}}}
    \;,\\
    C'&= \frac{4}{3} \left[-\frac{5 d^{R_F} C_2^{\text{adj}} C_2^{R_F}
   N_F}{d^{\text{adj}}}+\frac{3 d^{R_F} \left(C_2^{R_F}\right)^2 N_F \left(10
   N_F-d^{R_F}\right)}{d^{\text{adj}} \left(2 N_F+d^{R_F}\right)}+17
   \left(C_2^{\text{adj}}\right)^2\right]
   \;.
\end{align}
Here, $C_2^\text{adj}$ and $d^\text{adj}$ refer to the second Casimir and the dimension of the adjoint representation, respectively. Similarly, $C_2^{R_F}$ and $d^{R_F}$ denote the same for the fermionic representation\footnote{Either of the latter group-theoretic invariants can be traded for the Dynkin index 
\begin{align}
    S^{R_F} \equiv \frac{d^{R_F}}{d^\text{adj}}\,C_2^{R_F}
\end{align}
but note that the latter is defined only up to a constant and varying conventions are used in the literature. Here, we use the dimension and the second Casimir.}.

Naively, there are two ways to achieve perturbativity of the possible fixed points
$\alpha_{\ast,\,\text{BZ}}=-\frac{B}{C}$
and
$\alpha_{\ast,\,\text{GY}}=-\frac{B}{C'}$, i.e., either by (i) making $B$ small, or by (ii) making $C$ or $C'$ large. It is typically not possible to achieve the latter (as a function of $N_C$ and $N_F$ for instance) without invalidating perturbation theory at higher orders. The subsequent discussion of the $U(1)$ in Sec.~\ref{sec:triviality} will serve as an explicit example. On the contrary, non-Abelian gauge groups can allow for perturbatively small $B$ without invalidating perturbation theory \cite{Litim:2014uca}.
A dedicated 3-loop analysis of a simple $SU(N)$ gauge group with fermions in the fundamental representation \cite{Bond:2017tbw} provides strong indications that perturbative yet interacting gauge-Yukawa fixed points are only possible for $N_C\geqslant 5$. However, this does not necessarily imply that the same conclusions hold for arbitrary representations. For extensions of the SM, this has been tested by an explicit grid search for a single type of BSM representation in~\cite{Barducci:2018ysr}. Before extending such a grid search to multiple different types of BSM representation, we discuss each of the simple SM subgroups on its own. This provides a good intuition of why certain phases, cf.~Fig.~\ref{fig:phaseStructure}, are possible and others are not.

\subsection{The non-Abelian subgroups of the SM}

\begin{figure}
    \centering
    \includegraphics[width=0.243\textwidth]{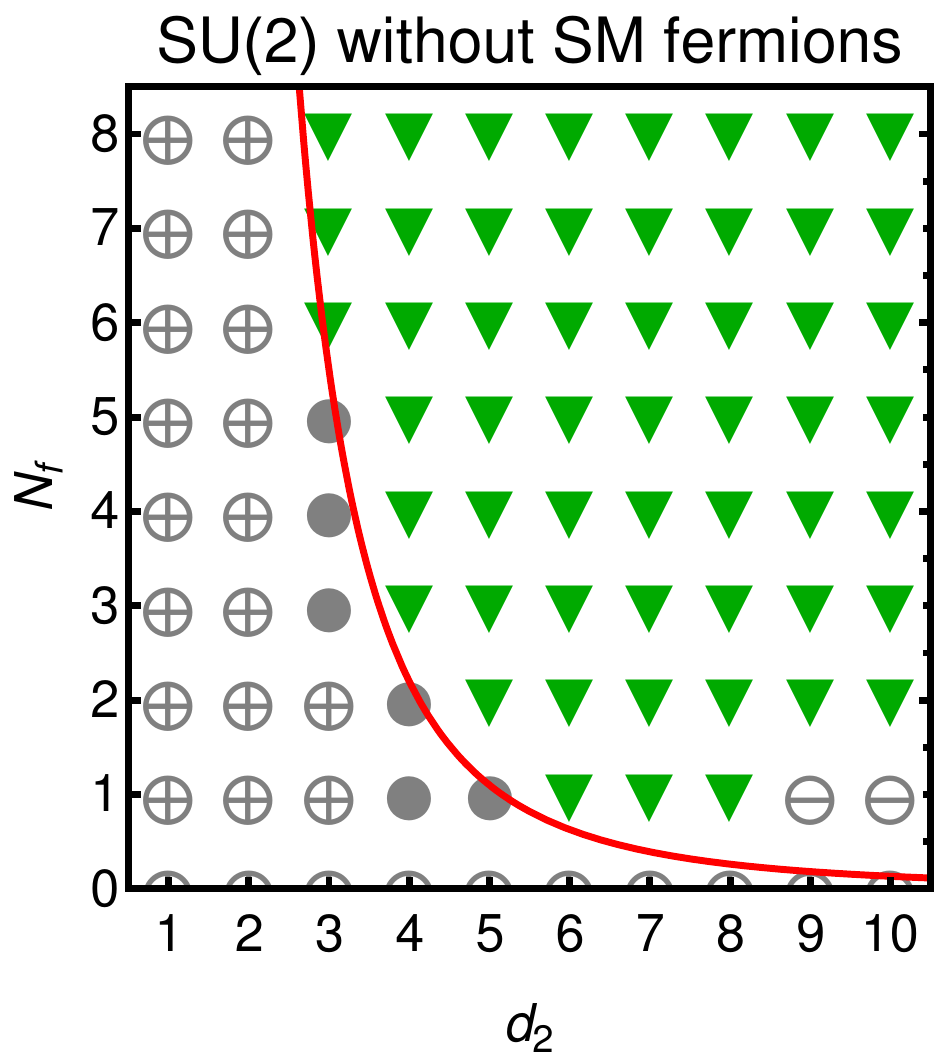}
    \includegraphics[width=0.243\textwidth]{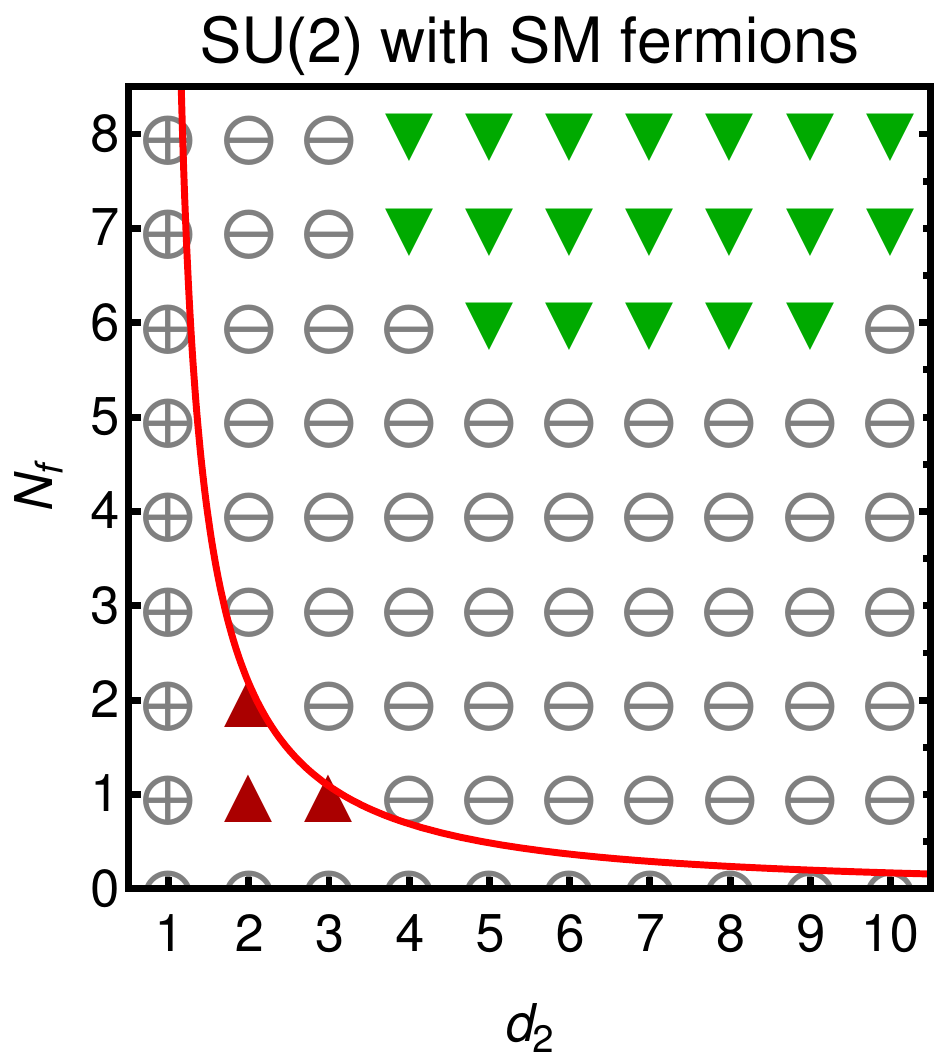}
    \vline
    \includegraphics[width=0.243\textwidth]{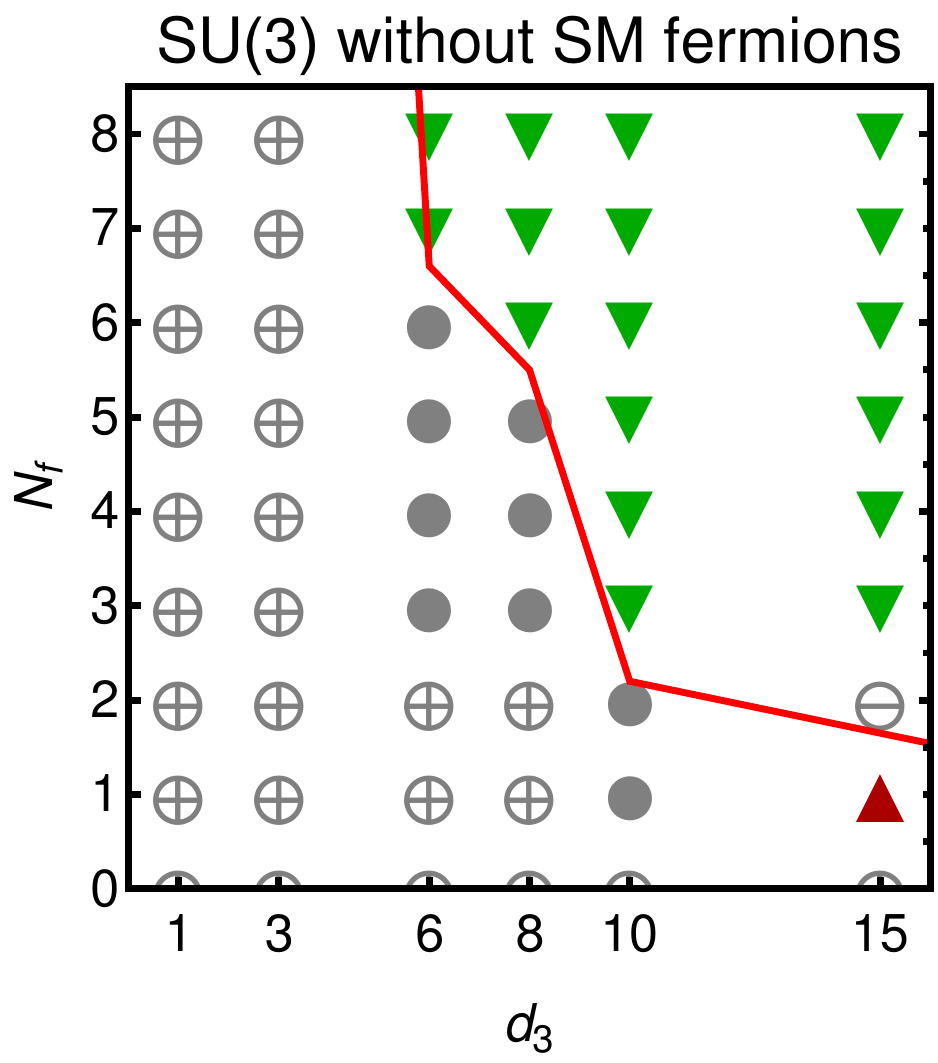}
    \includegraphics[width=0.243\textwidth]{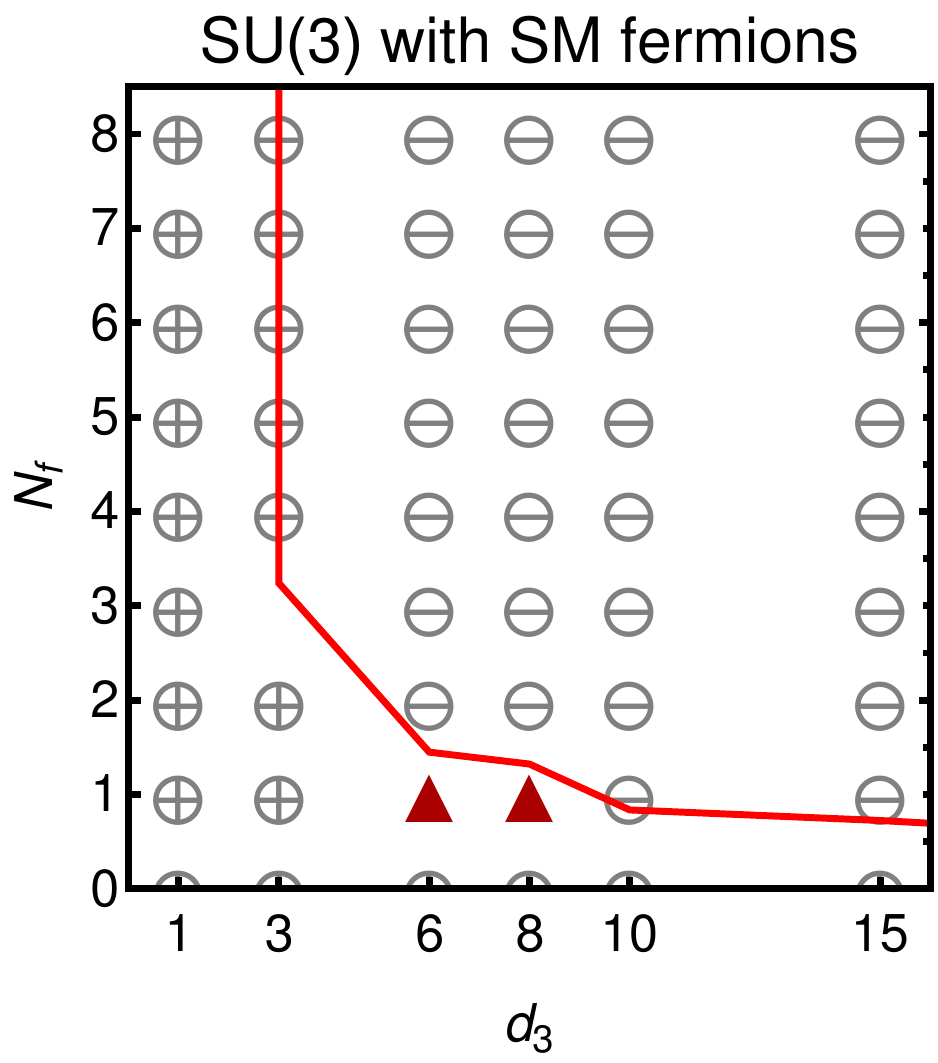}
\caption{
\label{fig:accessibleFPs}
Different phases, i.e., LS ($\blacktriangledown$), GY ($\blacktriangle$), BZ ($\CIRCLE$), CAF ($\oplus$), and CT ($\ominus$), of SU(2) and SU(3) gauge groups depending on the number and dimension (cf.~Eqs.~\eqref{eq:groupTheoryData_SU2}-\eqref{eq:groupTheoryData_SU3}) of the included BSM representations and on whether the SM matter fields are included or not. The thick red curve highlights where the 1-loop contribution vanishes, i.e., $B=0$. Interacting fixed points can only be controlled perturbatively if they lie in the vicinity of this line. The LS phase recedes from the perturbatively accessible region whenever the SM fields, i.e., charged matter without or with neglibily small, Yukawa couplings are included. 
}
\end{figure}

Which of the gauge-Yukawa phases is accessible in perturbation theory depends on the sign of $C'$ in the region close to a sign-change of $B$, cf.~Fig.~\ref{fig:phaseStructure}. Note that the sign of $C$ is always fixed close to a sign change of $B$, cf.~\cite{Bond:2016dvk}. 
The sign of $C'$, in turn, depends on the specific gauge group and matter representations. 
In particular, additional fermionic representations without (or with negligibly small) Yukawa couplings result in additional screening contributions to $B$ and $C$, while they do not contribute to $(C'-C)$ since they do not participate in Yukawa interactions. Hence, charged fermions without Yukawa couplings will influence which phases are available. 

The latter also occurs in the SM where there are 32 light Weyl degrees of freedom with negligibly small Yukawa couplings\footnote{This counting excludes the top quark since its Yukawa coupling is not negligibly small. It also excludes potential right-handed neutrinos which are SM singlets anyway.}. We explicitly visualize their significance for the existence of gauge-Yukawa fixed points in the case of SU(2) and SU(3) in Fig.~\ref{fig:accessibleFPs}. 
Without the SM fermions, there exist BSM representations (such as the $d_2=3$ dimensional for SU(2) and the $d_3=8$ dimensional for SU(3)) for which, with growing number $N_F$ of BSM fermions, $B$ changes sign before $C'$ does. As a function of $N_F$ one moves from complete asymptotic freedom to the Banks-Zaks phase and into the Litim-Sannino phase, i.e., through the chain $\text{CAF}\rightarrow\text{BZ}\rightarrow\text{LS}$, cf.~first and third panel in Fig.~\ref{fig:accessibleFPs}. In particular, one enters the LS phase via a sign change in $B$, i.e., in the region in which the interacting fixed points can be perturbatively controlled.
Inclusion of the SM fermions prohibits the realization of this chain, i.e., there is no possible BSM representation for which $B$ changes sign before $C'$ does. When adding additional BSM representations one therefore always follows a different chain with growing $N_F$: starting from complete asymptotic freedom and moving through the Banks-Zaks phase, one instead enters the gauge-Yukawa and ends up in the completely trivial phase, i.e., this realizes the chain $\text{CAF}\rightarrow\text{BZ}\rightarrow\text{GY}\rightarrow\text{CT}$. For most BSM representations which can be added to the SM case, the BZ and GY phase only occur at non-integer values of $N_F$ such that this formal chain is effectively reduced to $\text{CAF}\rightarrow\text{GY}\rightarrow\text{CT}$ or $\text{CAF}\rightarrow\text{CT}$, cf.~Fig.~\ref{fig:accessibleFPs}. Formally, this chain can be prolonged and the LS phase can still be entered from the CT phase, cf.~upper-right area of the second panel in Fig.~\ref{fig:accessibleFPs}. The minimal (but quite large) number of BSM fermions identified in \cite{Bond:2017wut} realizes this formal window of the LS phase. While this can occur at small values of the couplings if $C'\gg B\gg 1$, the latter invalidates perturbation theory and such fixed points are lost at NNLO, cf.~also \cite{Barducci:2018ysr}.

Regarding extensions of the SM, we can conclude that the SM fermions with negligibly small Yukawa couplings prohibit from entering the LS phase, i.e., no perturbatively controlled, interacting fixed points with UV-attractive directions are possible. On the contrary, fully IR-attractive interacting gauge-Yukawa fixed points in the GY phase remain possible for special dimension and number of BSM representations, cf.~red upward triangles in Fig.~\ref{fig:accessibleFPs}.
As we shall see in Sec.~\ref{sec:newMatter}, both conclusions persist for the full SM gauge group. The GY phase realizes \emph{effective asymptotic safety} if the theory space is extended to include mass terms or scalar vacuum expectation values. In this case, the theory departs from (close to) the otherwise fully IR-attractive fixed point at RG scales below this mass threshold.

\subsection{Persistence of Abelian triviality}
\label{sec:triviality}
Despite the complete asymptotic freedom of both non-Abelian subgroups, the SM is not UV-complete, i.e., it eventually breaks down at a transplanckian but finite energy scale. Due to the lack of antiscreening self-interactions in the U(1) gauge group, matter fluctuations dominate and screen the associated Abelian gauge coupling. At $\sim 10^{41}$ GeV, the latter grows beyond perturbative control and eventually results in a perturbative divergence -- the Landau pole \cite{GellMann:1954fq}. Beyond perturbation theory, the U(1) triviality problem has been confirmed by different non-perturbative methods \cite{Gockeler:1997dn, Gockeler:1997kt, Gies:2004hy}, but so far only in the absence of Yukawa couplings.

Indeed, the presence of a Yukawa coupling formally places an Abelian gauge group in the Litim-Sannino phase, cf.~Section~\ref{sec:RGstructure}.
Unfortunately, the corresponding interacting pseudo-fixed-point cannot occur within the perturbatively controlled regime. Since we found no explicit discussion of the latter statement in the literature, we will provide it in the following. 

In principle, every U(1) gauge group with $N_F$ fermions of charge $Y$ and associated scalars to facilitate Yukawa couplings is in the LS phase which would indicate the presence of an interacting UV fixed point for the gauge coupling at
\begin{align}
\label{eq:fidU1LSFP}
    \alpha_{U(1)\,\ast} = \frac{1}{15\,Y^2}\;,
    \quad\quad\quad
    \alpha_{y\,\ast} = \frac{2}{5\,N_F}\;.
\end{align}
The explicit NLO and NNLO $\beta$-functions are presented in App.~\ref{app:U1coeffs}.
It would seem as if this fixed point becomes more perturbative for large $Y^2$ and large $N_F$ but this ignores the accompanying growth of higher-loop contributions and the resulting breakdown of perturbation theory. To properly analyze the above fixed point, one has to introduce a t'Hooft-like rescaling of the couplings. More specifically, one has to rescale the couplings $\alpha_g$ and $\alpha_y$ such that all higher-loop contributions either vanish or at least converge to finite values at large $N_F$ and large $Y^2$. In the present case, the minimal rescaling that suppresses all higher-loop contributions with growing $N_F$ and $Y^2$ is given by
\begin{align}
    \alpha_{U(1)} = \frac{\widetilde{\alpha}_{U(1)}}{N_F\,Y^2}\;,
    \quad\quad\quad
    \alpha_y = \frac{\widetilde{\alpha}_y}{N_F}\;.
\end{align}
The correspondingly rescaled $\beta$-functions reveal that (in contrast to the non-Abelian case in \cite{Litim:2014uca}) only the trivial fixed point persists in the perturbative large-charge--large-$N_F$ limit.
We have explicitly confirmed that for \emph{any} combination of integer $N_F\geqslant 0$ and arbitrary $Y^2$, the absolute value of the NNLO contributions is larger than that of the NLO contributions when evaluated at the fiducial fixed point in Eq.~\eqref{eq:fidU1LSFP} -- a clear sign that perturbation theory is not valid anymore.

The physical mechanism through which the Litim-Sannino fixed point arises, i.e., the balance of screening contributions from fermionic fluctuations against antiscreening contributions from Yukawa couplings, is present nevertheless. Thus, it might be worthwhile to conduct a non-perturbative analysis of this fixed-point mechanism in Abelian theories with Yukawa couplings in the future. However, for the present perturbative analysis, we conclude that the Abelian gauge group of the SM will always remain trivial.

\section{A quantitative measure of predictivity}
\label{sec:perturbativityCriterion}

For a given gauge-Yukawa theory with fixed gauge group and matter content, we define the perturbative range of coupling values $\alpha_i$ by the condition that all NNLO contributions remain smaller than the respective NLO contributions, i.e., 
\begin{align}
\label{eq:perturbativityCriterion}
    \text{perturbativity}
    \quad\Leftrightarrow\quad    
    |\mathbf{\beta}^{(NNLO)}|<\frac{1}{2}|\mathbf{\beta}^{(NLO)}|\;.
\end{align}
The factor $\frac{1}{2}$ is included such as to avoid the regime of novel fiducial fixed points arising at NNLO. Another reason for the inclusion of this factor is the U(1) Landau pole, as will become clear below.
This perturbativity condition is rather non-conservative, meaning that perturbation theory may break down earlier.
The resulting set of perturbative EFTs encloses a finite volume $\mathcal{V}$ in the (truncated) theory space of all couplings\footnote{Since we work in the perturbative regime, all higher-order couplings will necessarily remain irrelevant. Hence, the UV-complete region does not extend in any of these directions and its volume, if finite in truncated theory space, remains finite in full theory space. Technically, this is not necessarily true for the overall EFT volume in the theory-space volume which permits an extension in any higher-order direction of the full theory space. We restrict to truncated theory space in the following.}. More explicitly, we use the volume of the convex hull obtained from a Delauney-triangulation of a large enough random set of points in theory space which fulfill the perturbativity criterion\footnote{One can easily see that the convex hull is not always a good approximation to the theory-space volume enclosed by the separatrices between fixed points, cf.~upper right-hand panel in Figure~\ref{fig:phaseStructure}. However, it is (to our knowledge) the only mathematically well-defined discrete notion of such a volume. It certainly suffices to quantify the statements of this study.}. We ensure convergence of this discrete predictivity measure by averaging over several individual random sets of perturbative EFTs and making sure that the statistical error is subleading. 

This allows us to define a quantitative measure of predictivity. The theory-space volume $\mathcal{V}$ can be evolved by following the RG flow to the IR. Given the initial volume $\mathcal{V}_\Lambda$ at the cutoff scale $\Lambda$, and its evolution following the RG flow, i.e., $\mathcal{V}_k$, at RG scale $k$, we define predictivity $\mathcal{P}(k)$ by
\begin{align}
    \label{eq:predictivityMeasure}
    \mathcal{P}(k) = \frac{\mathcal{V}_k}{\mathcal{V}_\Lambda}\;.
\end{align}
We call EFTs predictive (non-predictive), whenever their theory-space volume decreases (increases) along the flow. 
Non-predictive EFTs tend to formally result in $\mathcal{P}(k)\rightarrow\infty$ at finite $k<\Lambda$ which signals that they have diverged beyond perturbative control.
In the predictive case, however, $\mathcal{P}(k)$ provides a quantitative measure of how predictive the EFT is.

It will also prove useful to exclude specific couplings, e.g., the measured SM couplings, from the predictivity measure and instead match them to their experimentally known values at a specified low-energy scale, e.g., at $\Lambda_\text{ew}<\Lambda$. The resulting $\tilde{\mathcal{P}}(k)$ can measure partial predictivity, even if the overall EFT is classified as non-predictive.  

Both, the predictivity and the partial predictivity measure do not necessarily rely on perturbation theory and can be applied to (sufficiently converged) non-perturbative truncations of theory space as well. However, they do require to define an initial volume in theory space in which the present truncation is sufficiently converged, i.e., a non-perturbative analog of Eq.~\eqref{eq:perturbativityCriterion}. Whenever such an initial volume in theory space can be defined, its evolution under the RG flow allows us to quantify the predictivity of \emph{effective asymptotic safety} via the measure in Eq.~\eqref{eq:predictivityMeasure}. In particular, this applies to truncations of the Reuter universality class \cite{Reuter:1996cp}, see~\cite{Percacci:2017fkn,Reuter:2019byg, Eichhorn:2020mte} for introductory texts and \cite{Percacci:2010af,Eichhorn:2017eht,Eichhorn:2018yfc,deAlwis:2019aud} for previous discussions in the \emph{effective asymptotic safety} context. We leave such an analysis for future work.
To exemplify these definitions, we will discuss the heavy gauge-Yukawa sector of the SM in Section~\ref{sec:SM} before adding new matter degrees of freedom in Section~\ref{sec:newMatter}.

\section{The heavy-top limit of the Standard Model}
\label{sec:SM}
We focus on the heavy gauge-Yukawa sector of the SM, i.e., on the three gauge couplings $\alpha_{1,\,2,\,3}$ and the top-Yukawa coupling $\alpha_t$. It is a very good approximation to assume all other fermions as being massless, i.e., to set their Yukawa couplings to zero.
Similarly, we neglect contributions from the quartic coupling $\lambda_4$ which is also negligible with regards to the gauge-Yukawa sector, as long as all couplings remain within the perturbative regime because it only arises at 2-loop and 3-loop order for Yukawa and gauge couplings, respectively. Supplementary conditions implied by stability conditions of the Higgs potential \cite{Litim:2015iea, Bond:2017tbw} are deferred to future studies.
\begin{figure}
    \centering
    \includegraphics[height=0.34\textwidth]{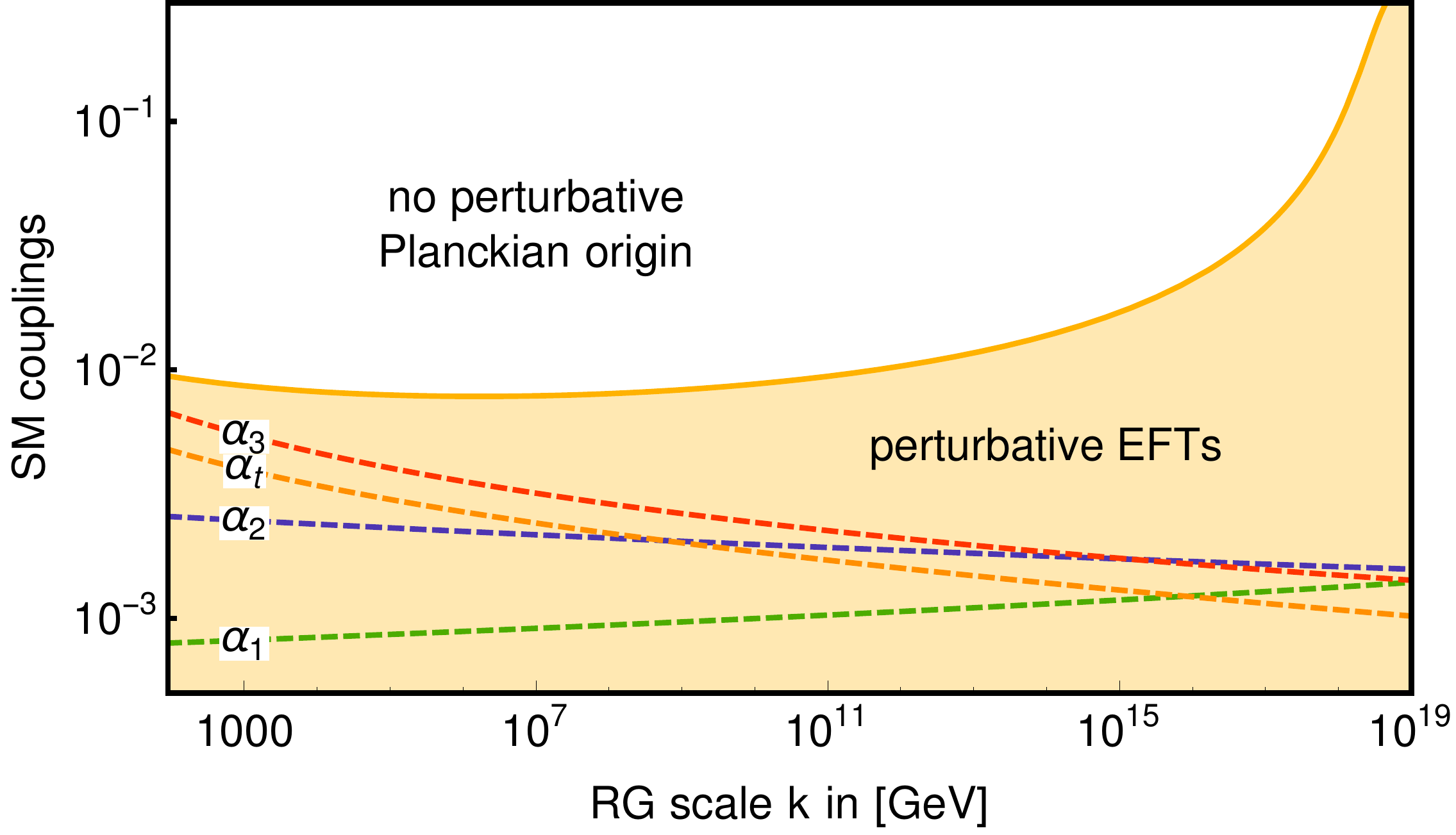}
    \hspace*{20pt}
    \includegraphics[height=0.34\textwidth]{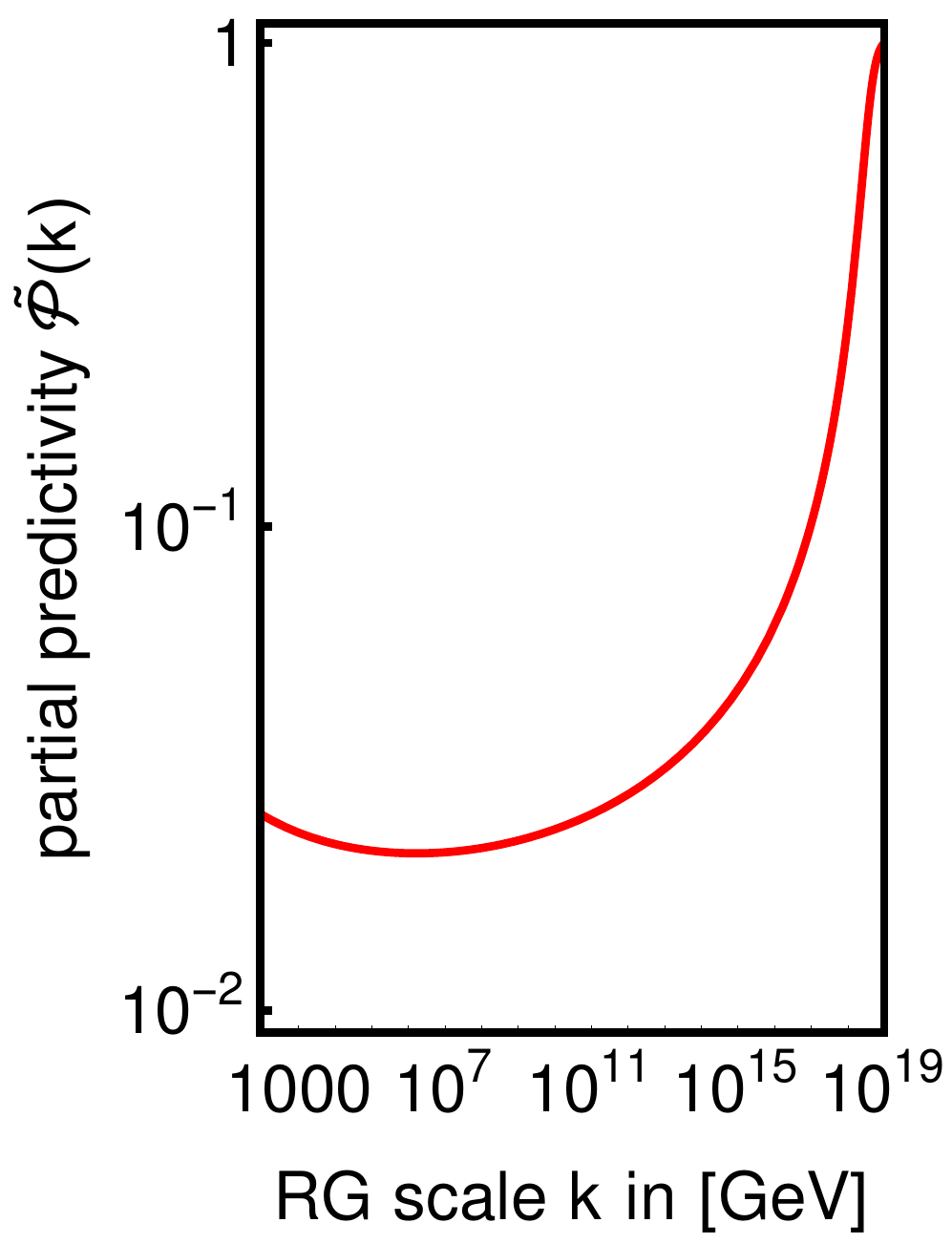}
    \hspace*{20pt}
\caption{
\label{fig:topPredictivity}
Left-hand panel: RG flow of the SM gauge-Yukawa theory. The shaded region indicates values for $\alpha_t$ which can originate from a perturbative EFT at the cutoff scale $\Lambda_\text{Planck}$. The focusing of this region towards the lower scales exemplifies the partial predictive power of the SM as an EFT. The dashed trajectories indicate the RG flow of $\alpha_3$, $\alpha_2$, $\alpha_1$, and $\alpha_t$, matching observed values at the electroweak scale.
Right-hand panel: Evolution of the partial predictivity measure with the RG flow.
}
\end{figure}
\subsection{Partial predictivity within the Standard Model}

The heavy SM is non-predictive, i.e., $\mathcal{P}(k)$ quickly diverges below the cutoff scale. This is a result of the antiscreening nature of the non-Abelian gauge couplings realizing the CAF phase (cf.~Section~\ref{sec:RGstructure}) in the SM. Coupling values at the cutoff scale that lie close to the edge of the perturbative regime will quickly be driven to values beyond perturbative control towards the IR.

On the other hand, if one excludes the non-Abelian gauge couplings from the predictivity measure and instead fixes them to their known experimental values at the electroweak scale, the SM is partially predictive in the remaining theory space. This is a consequence of the screening nature of both the top-Yukawa and the U(1) gauge coupling. When excluding also the U(1) gauge coupling from the predictivity measure, the resulting partial predictivity $\tilde{\mathcal{P}}(k)$ reflects the pre-Tevatron situation in which all the gauge couplings had already been experimentally measured, while the top-Yukawa coupling $\alpha_t$ remained unknown. Figure~\ref{fig:topPredictivity} shows the evolution of $\tilde{\mathcal{P}}(k)$ along the RG flow. In this simple one-dimensional slice of theory space, the predictivity measure simply amounts to the normalized evolution of the full perturbative range of top-Yukawa values below $\Lambda_\text{Planck}$. Hence, enforcing a perturbative origin at $\Lambda_\text{Planck}$ bounds the top quark to be lighter than $M_t\lesssim 210\,\text{GeV}$. The underlying reason is the associated partial IR fixed point for Yukawa couplings in gauge-Yukawa theories previously uncovered in~\cite{Pendleton:1980as,Hill:1980sq,Wetterich:1981ir}, cf. also Eq.~\eqref{eq:partialYukawaFP}.

\subsection{The Landau pole remains transplanckian}
As discussed in Sec~\ref{sec:triviality}, the triviality of the $U(1)$ hypercharge cannot be cured within perturbation theory. On the other hand, the associated Landau pole remains above the Planck scale as long as the other SM couplings remain within the perturbative regime (and no BSM representations with hypercharges are added).

Even in the absence of any new states with hypercharge, NLO and NNLO contributions from the non-Abelian gauge and top-Yukawa couplings in a modified BSM RG flow can potentially further screen the U(1) gauge coupling and therefore result in a lowered Landau pole, cf.~also~\cite{Bond:2017wut}. However, for any perturbative extension of the SM that still matches the measured electroweak-scale value for $\alpha_1$, the Landau pole remains at transplanckian energies. One can numerically determine that
$\alpha_3\lesssim 0.15$, 
$\alpha_2\lesssim 0.09$, and 
$\alpha_t\lesssim 0.53$
is required to conform to the perturbativity criterion in Eq.~\eqref{eq:perturbativityCriterion}, i.e., to $|\mathbf{\beta}^{(NNLO)}|<\frac{1}{2}|\mathbf{\beta}^{(NLO)}|$. These maximal values have been determined by a grid search at random $\alpha_1$. We then fix the non-Abelian gauge couplings and the top Yukawa coupling to these maximal values. By definition, any RG flow within the perturbative regime cannot outgrow these values. Numerical integration of the resulting RG flow of the U(1) coupling shows that the U(1)-Landau pole remains safely beyond the Planck scale.

\begin{figure}
    \centering
    \includegraphics[height=0.35\textwidth]{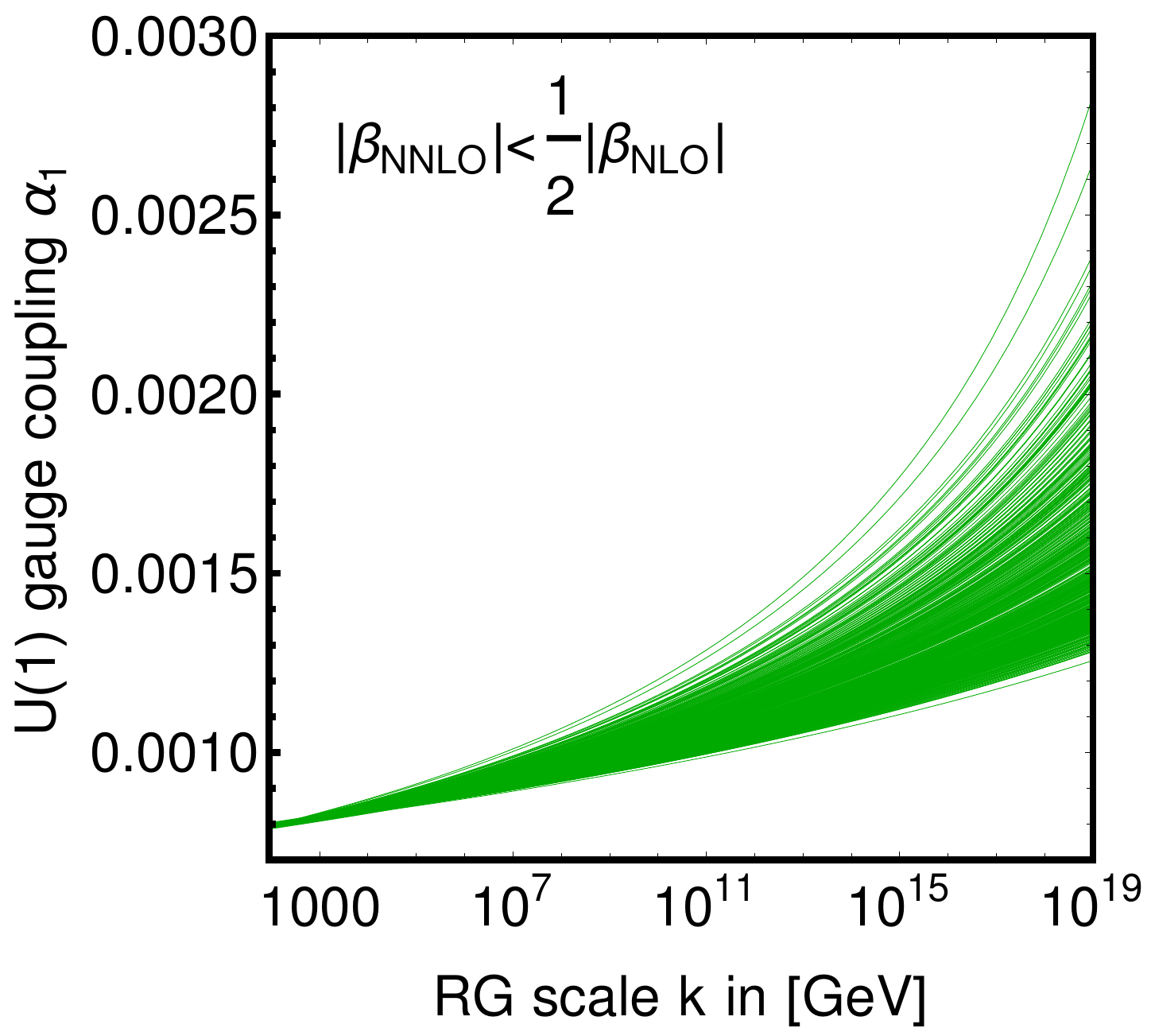}
    \hspace*{20pt}
    \includegraphics[height=0.35\textwidth]{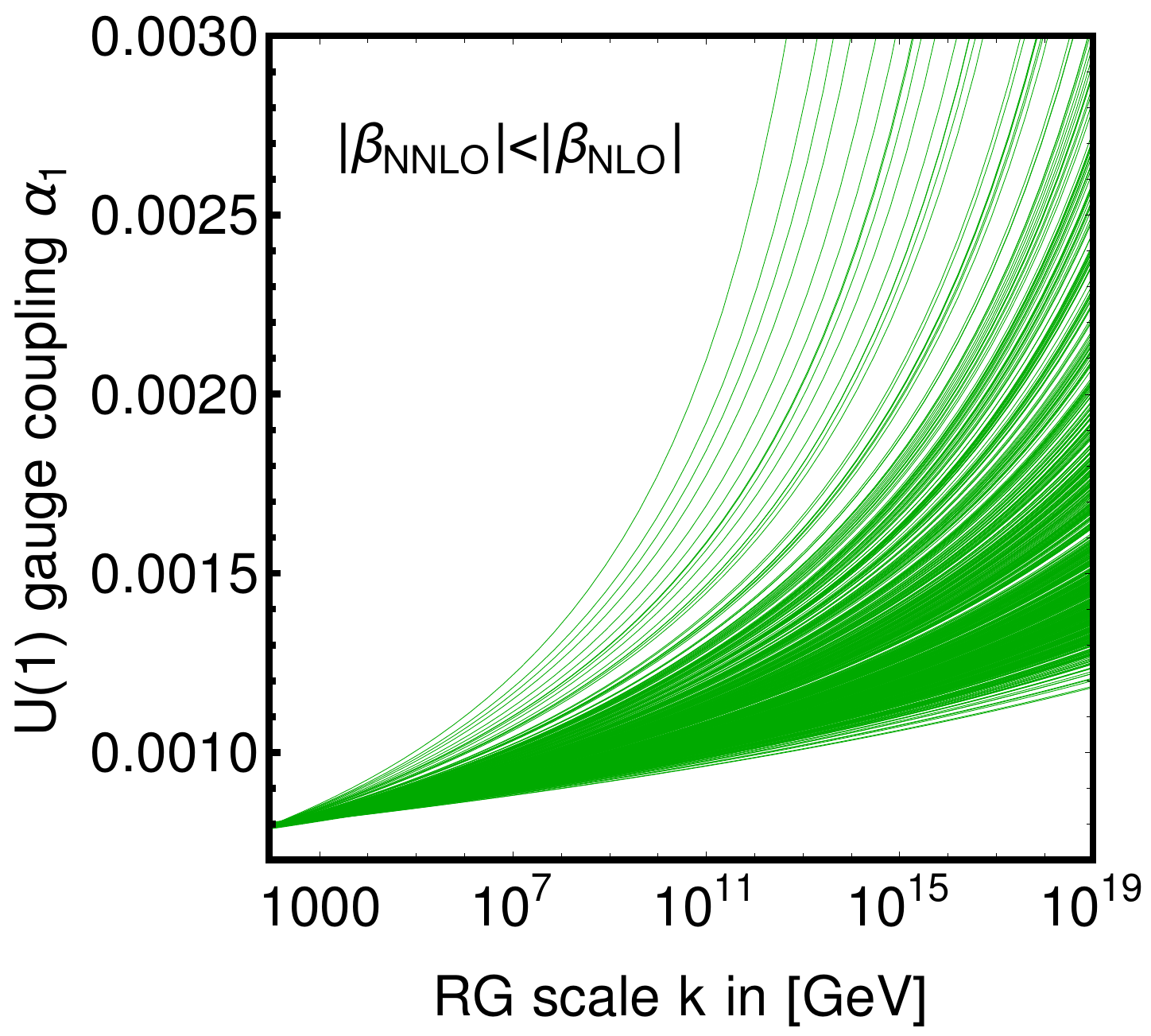}
    \hspace*{20pt}
\caption{
\label{fig:landauPole}
RG-flow of the U(1) gauge coupling matched to the observed electroweak value for arbitrary non-Abelian gauge and top-Yukawa couplings within the perturbative regime. Left-hand panel: perturbativity defined by $|\mathbf{\beta}^{(NNLO)}|<\frac{1}{2}|\mathbf{\beta}^{(NLO)}|$, cf.~Eq.~\eqref{eq:perturbativityCriterion}. Right-hand panel: perturbativity defined by $|\mathbf{\beta}^{(NNLO)}|<\frac{1}{2}|\mathbf{\beta}^{(NLO)}|$.
}
\end{figure}

The left-hand panel in Figure~\ref{fig:landauPole} shows the RG-flow of the Abelian gauge coupling matching to the observed electroweak-scale value for a random set of fixed values of the other heavy-SM couplings satisfying the perturbativity criterion. Subplanckian Landau poles are not present. Loosening the perturbativity criterion in Eq.~\eqref{eq:perturbativityCriterion} to $|\mathbf{\beta}^{(NNLO)}|<|\mathbf{\beta}^{(NLO)}|$ allows for rare cases at the edge of the redefined perturbative regime for which the Landau pole is shifted slightly below the Planck scale, cf.~right-hand panel in Figure~\ref{fig:landauPole}. In any case, all the perturbative BSM fixed points discussed in Sec.~\ref{sec:newMatter} are much more perturbative than any of the above bounds.

We conclude that the persistence of a U(1)-Landau pole -- at least in any of the subsequently important BSM scenarios in which the BSM representations do not carry hypercharge -- is no meaningful criterion in the search for physically interesting fixed points in the framework of perturbative EFTs below the Planck scale. Instead, one should merely verify that the Landau pole remains transplanckian. In this aspect, we advocate a different point of view, than, e.g.,~\cite{Barducci:2018ysr}.

\section{New matter degrees of freedom}
\label{sec:newMatter}

In the following, we allow for additional fermionic matter in arbitrary representations (as well as for the associated uncharged scalars to facilitate Yukawa couplings). 
We have seen that, within perturbation theory, any $U(1)$ factor will remain trivial. Therefore, we do not attempt to modify the RG flow of the $U(1)$ gauge coupling and thus only add BSM fermions which are uncharged under the $U(1)$. We allow for an arbitrary number of different representations of BSM fermions, i.e., $N^{R_a}_F$ fermions in the $(d_2^{R_a},d_3^{R_a})$-dimensional representation of $SU(2)$ and $SU(3)$, respectively. The Lagrangian and the $\beta$-functions for the three gauge couplings, the top-Yukawa coupling, as well as additional BSM Yukawa couplings, i.e., for
\begin{align}
\alpha_1 = \frac{g_1^2}{(4\pi)^2}\;,\quad
\alpha_2 = \frac{g_2^2}{(4\pi)^2}\;,\quad
\alpha_3 = \frac{g_3^2}{(4\pi)^2}\;,\quad
\alpha_{y_{t}} = \frac{y_{t}^2}{(4\pi)^2}\;,\quad \alpha_{y}^{R_a} = \frac{y_{R_a}^2}{(4\pi)^2}\;,
\end{align}
are generalized from \cite{Barducci:2018ysr} and collected in App.~\ref{app:betasForBSM}.

With the intuition from the results in Sec.~\ref{sec:RGstructure} for simple non-Abelian gauge groups, we anticipate that, also in the semi-simple case, the non-Abelian subgroups cannot admit perturbatively controllable Litim-Sannino fixed points with an IR-repulsive (UV-attractive) direction. We confirm this expectation in the following explicit analysis. Fully IR-attractive gauge-Yukawa fixed points, on the other hand, can exist. From the viewpoint of effective asymptotic safety, these are the most predictive and in that sense most interesting fixed points, anyhow.

The larger the dimension of the BSM representations, the greater their screening effect on the 1-loop coefficient of the associated non-Abelian gauge coupling. Thus, there exists an upper dimension $d_{2,\,\text{crit}}^{R_a} = 4$ and $d_{3,\,\text{crit}}^{R_a} = 10$ beyond which even a single additional BSM representation will always push the associated non-Abelian SM gauge group into the completely trivial phase. Hence, the set of possible BSM representations for which perturbative non-vanishing gauge-Yukawa fixed points might exist is limited and easily tractable.
With the help of computer algebra \cite{Mathematica}, we simply scan through all possibilities and identify those for which the NLO beta-functions exhibit a fixed point with
\begin{align}
\label{eq:targetFP}
    \alpha_{2\,\ast} > 0\;,\quad
    \alpha_{3\,\ast} > 0\;,\quad
    \alpha_{t\,\ast} > 0\;,\quad
    \text{and}\quad
    \alpha^{R_a}_{y\,\ast} > 0\quad\forall\,a\;.
\end{align}
We subsequently test the perturbativity of each of the resulting fixed points by initializing a numerical root search in the NNLO beta-functions at the NLO fixed-point values. If the latter converges, we compare whether the signs of the critical exponents of the NLO and NNLO fixed points match. (If the root search does not converge, we discard the NLO fixed point.) Thereby we can identify perturbative fixed points for which NNLO corrections are subleading\footnote{
One might be able to construct more elaborate search algorithms and thereby potentially identify additional gauge-Yukawa BSM theories with perturbatively controlled interacting fixed points and we do not claim completeness.
}.

Irrespective of the specific representation $(d_2^{R_1},d_3^{R_1})$ and the number of copies $N^{R_1}_F$, we find that a single type of BSM representation $R_1$ is insufficient to generate a fixed point at which both $\alpha_{2\,\ast}\neq 0$ and $\alpha_{3\,\ast}\neq 0$. IR-attractive gauge-Yukawa fixed points at which only one of the non-Abelian gauge couplings is non-vanishing are available in perturbation theory and have been identified in \cite{Barducci:2018ysr}.

Proceeding to two different types of representations, i.e., $R_1$ and $R_2$, we are able to identify a single combination of BSM representations for which a fixed point as in Eq.~\eqref{eq:targetFP} is possible, i.e.,
\begin{align}
\label{eq:predictiveBSMcontent}
    N_{R_1}=1\;\text{copy of the}\;\;(d_2^{R_1},d_3^{R_1}) = (3,1)\,
    \quad\text{and}\quad
    N_{R_2}=2\;\text{copies of the}\;\;(d_2^{R_2},d_3^{R_2}) = (1,6)\;.
\end{align}
For this specific combination of BSM representations, both non-Abelian gauge groups are in the GY phase. Hence, all possible combinations of gauge-Yukawa fixed points exist.
In particular, this includes a fully IR-attractive fixed point at
\begin{align}
\label{eq:predictiveFP}
    \alpha_{1\,\ast} = 0\;,\;\;
    \alpha_{2\,\ast} \approx 0.0131\;,\;\;
    \alpha_{3\,\ast} \approx 0.0033\;,\;\;
    \alpha_{t\,\ast} \approx 0.0124\;,\;\;
    \alpha^{R_1}_{y\,\ast} \approx 0.0082\;,\;\;
    \alpha^{R_2}_{y\,\ast} \approx 0.0394\;.
\end{align}
The fixed point persists at NNLO order.

To summarize, we find that by adding suitable matter content to the SM, the non-Abelian gauge-Yukawa sector of the SM can transition from the CAF-phase to the GY-phase, and of course to the CT-phase. The explicit study supports that neither the BZ-phase nor the LS-phase is possible, cf.~Section \ref{sec:noGoForSMsubgroups}. Within perturbation theory, the U(1) always remains in the CT-phase.

\begin{figure}
    \centering
    \includegraphics[width=0.341\textwidth]{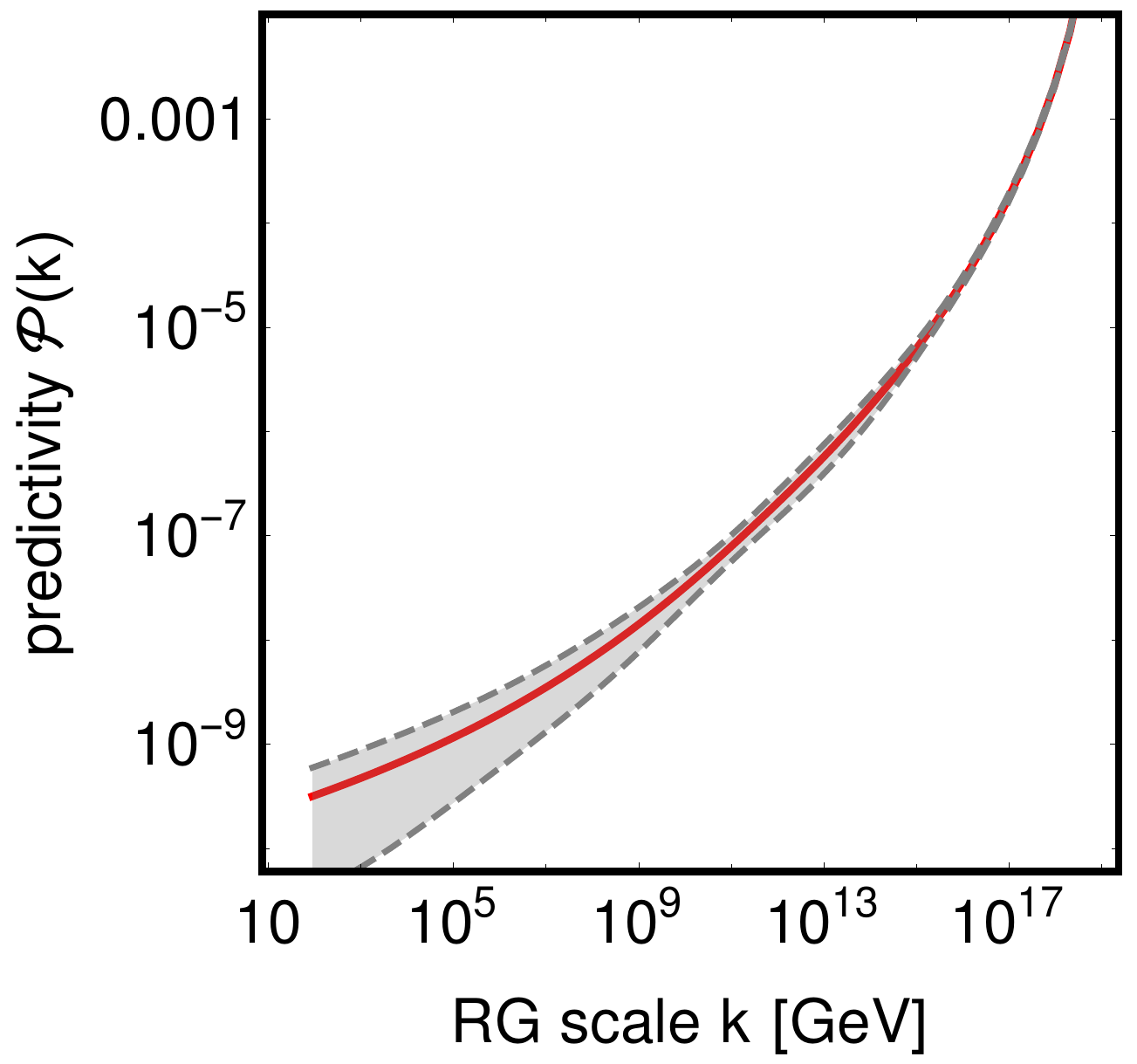}
    \hfill
    \includegraphics[width=0.31\textwidth]{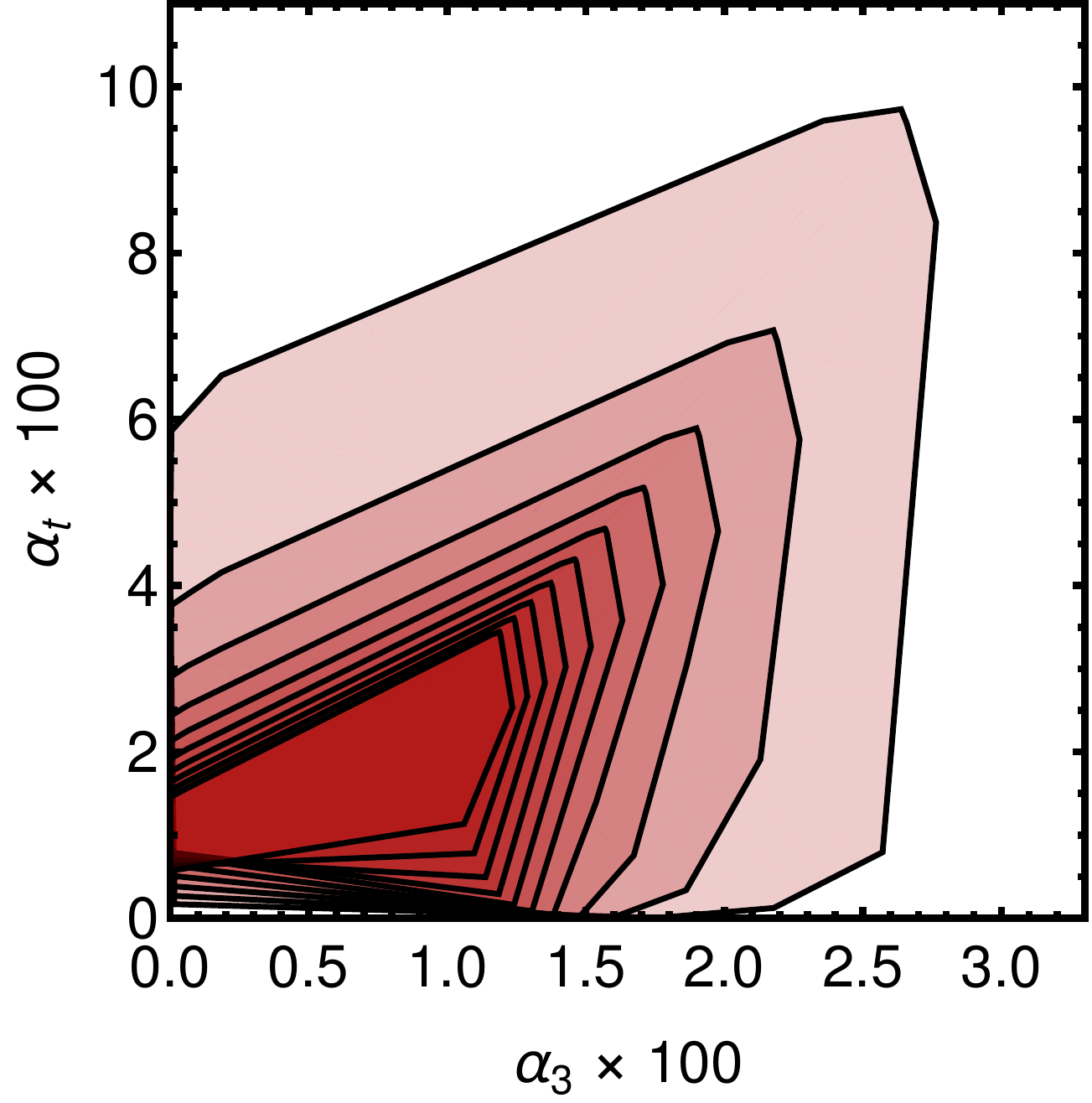}
    \hfill
    \includegraphics[width=0.312\textwidth]{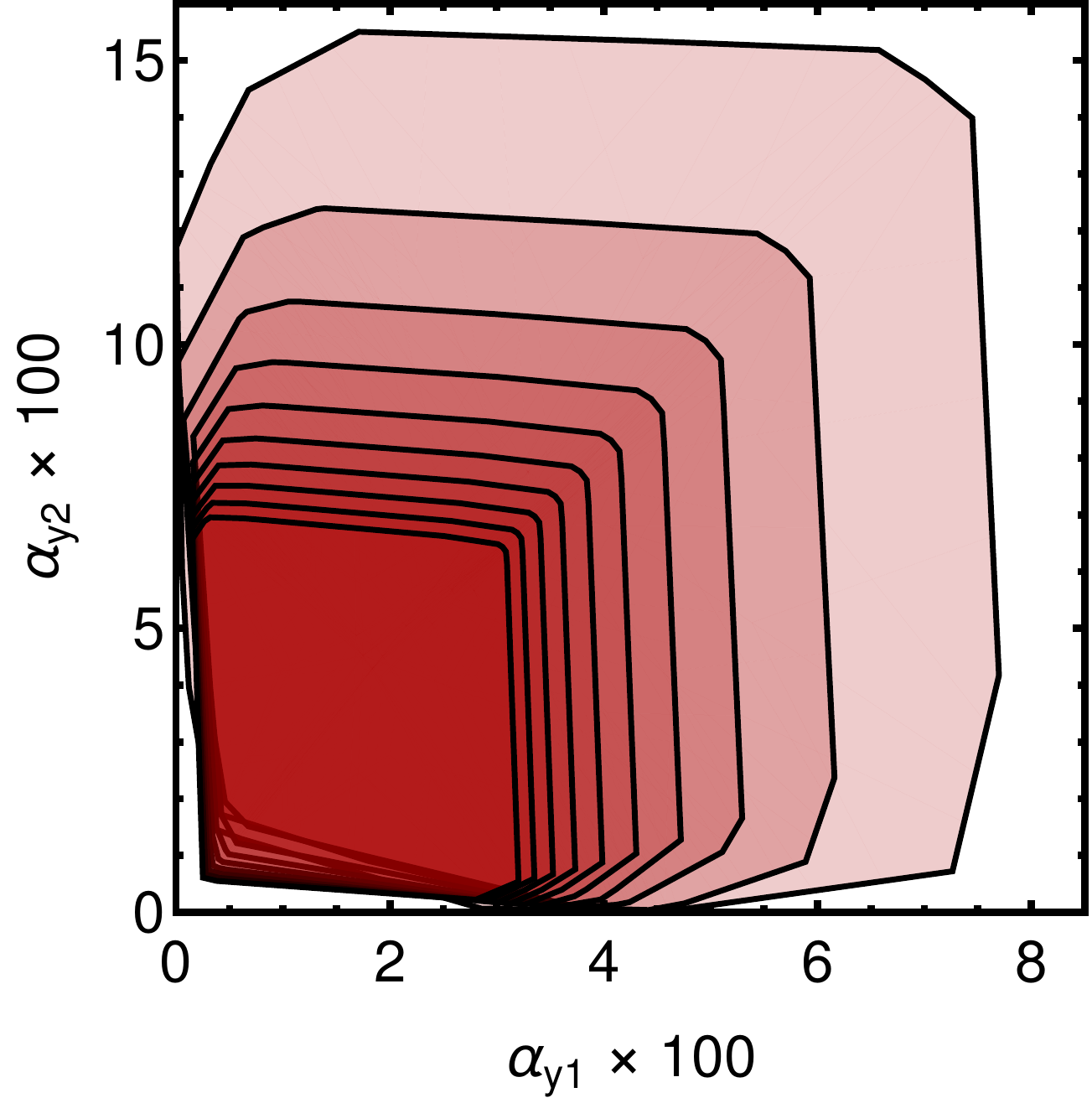}
\caption{
\label{fig:GY-phase_convexHullCollapse}
In the left-hand panel we show the predictivity of the BSM model identified in Eq.~\eqref{eq:predictiveBSMcontent}, averaged over 10 sets of perturbative but otherwise random initial EFTs. The gray-dashed region indicates the statistical error.
The other two panels show projections of the evolving theory-space volume (onto the $\alpha_3$-$\alpha_t$-plane (middle panel) and onto the $\alpha_y^{R_1}$-$\alpha_y^{R_2}$-plane (right-hand panel)). We plot its convex hull at each order of magnitude below $\Lambda_\text{Planck}$ with increasingly darker-red shading towards the IR.
}
\end{figure}
\subsection{Predictivity below the Planck scale}
For simple gauge-Yukawa theories in the CAF phase (BZ phase), the IR-complete region is reduced to the free theory (one-dimensional conformal window for vanishing Yukawa coupling), cf.~Figure~\ref{fig:phaseStructure}. Hence, these phases develop IR divergences for initial conditions that lie close to the edge of perturbativity at the cutoff scale. Put differently, they are non-predictive (as defined in Sec.~\ref{sec:perturbativityCriterion}).
On the contrary, the IR-complete region of theories in the GY or CT phase (and the LS) phase is two dimensional and covers all (or most) of the perturbative regime. Hence, these phases are predictive.

In Eq.~\ref{eq:predictiveBSMcontent}, we have identified a combination of BSM representations to push the non-Abelian SM subgroups into the predictive GY but not yet trivial phase. The two right-hand panels in Figure~\ref{fig:GY-phase_convexHullCollapse} depict the associated decreasing volume in theory space as a function of the RG-flow towards the IR in two slices of the overall 6-dimensional theory space. The left-hand panel shows the corresponding evolution of the predictivity measure $\mathcal{P}(k)$. Specifying to $\Lambda_\text{NP}=10^{5}\,\text{GeV}$, the theory-space volume is reduced by a factor of $\mathcal{P}(k=\Lambda_\text{NP})\sim 10^{-9}$ between $\Lambda_\text{Planck}$ and $\Lambda_\text{NP}$.

Despite fixed-point values that depart significantly, i.e., by several 100\%, from the measured SM values, predictivity is insufficient to exclude the BSM extension from matching to the SM electroweak scale. Put differently, the observed SM-coupling values lie within the `conformal' region of UV- and IR-complete theories (apart from the non-vanishing value of the Abelian gauge coupling, cf.~Sec.~\ref{sec:triviality}).

\subsection{Partial predictivity below the Planck scale}
\begin{figure}
    \centering
    \includegraphics[height=0.3\textwidth]{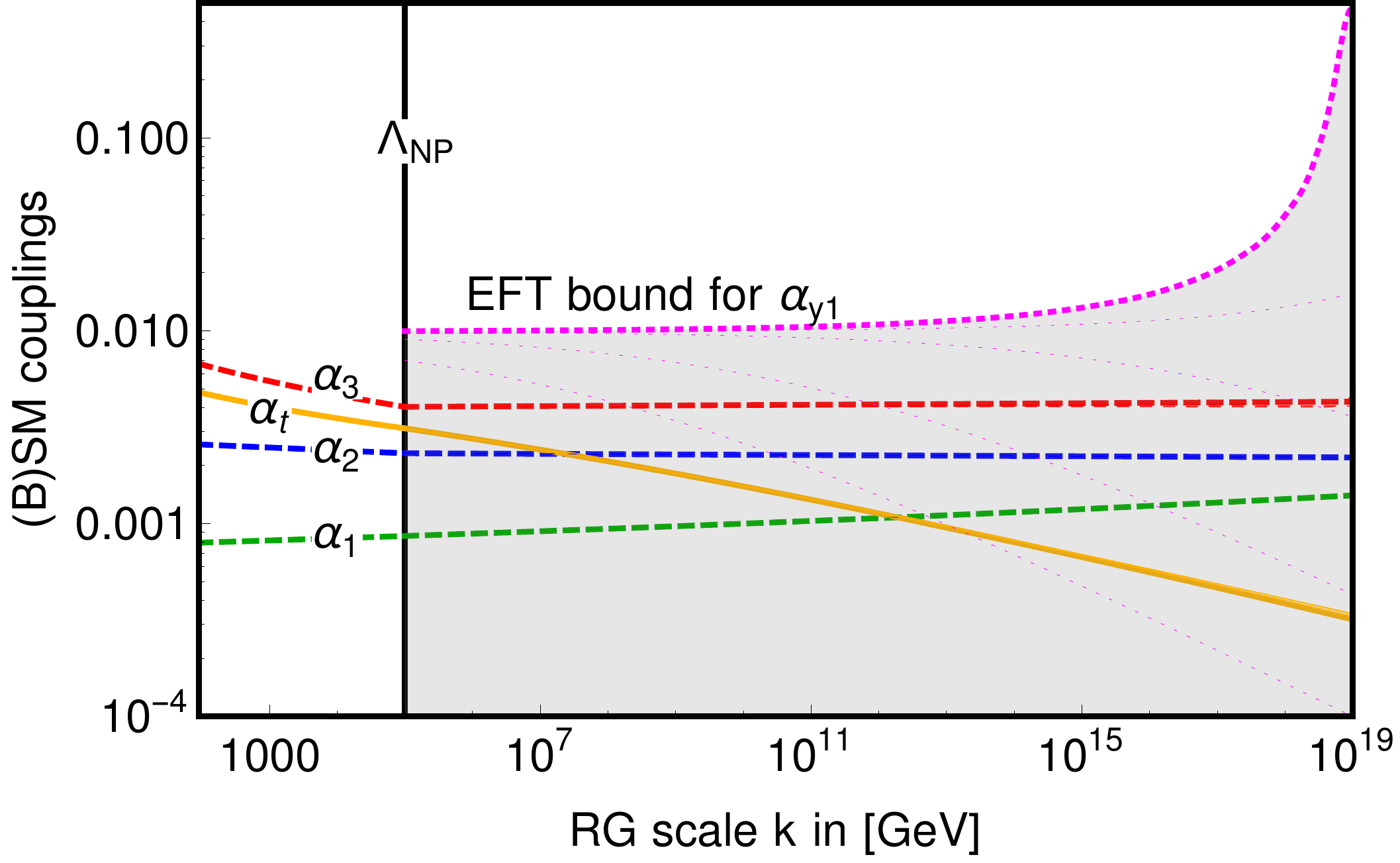}
    \hfill
    \includegraphics[height=0.3\textwidth]{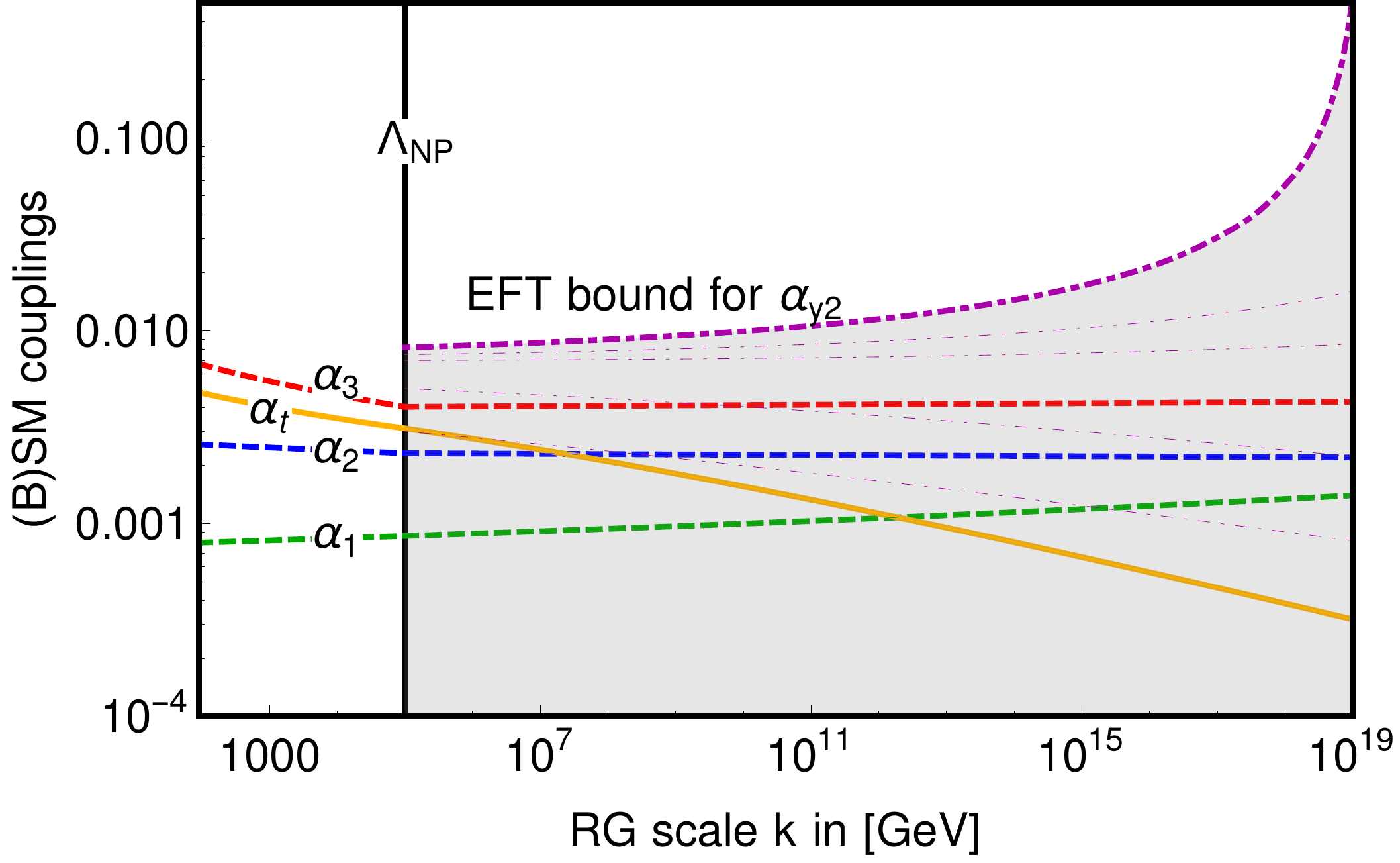}
\caption{
\label{fig:bsmPartialPredictivity}
Partial predictivity for the BSM theory identified in Eq.~\eqref{eq:predictiveBSMcontent}. The plots show RG trajectories that match the observed values of SM couplings in the heavy-top limit, i.e., $\alpha_1$, $\alpha_2$, $\alpha_3$ (dashed), and $\alpha_t$ (continuous). At energies below (above) $\Lambda_\text{NP}$, the BSM degrees of freedom decouple (are active). The BSM RG flow focuses arbitrary perturbative initial conditions for the BSM Yukawa couplings $\alpha_{y_1}$ (left-hand panel) and $\alpha_{y_2}$ (right-hand panel) at the Planck scale to the gray-shaded regions at lower scales. We also indicate (thin lines) several trajectories to exemplify the behavior of different RG trajectories within the conformal region.
}
\end{figure}
A phenomenologically more relevant question is that of partial predictivity under the condition of matching all the observed SM couplings, i.e., $\alpha_1$, $\alpha_2$, $\alpha_3$, and $\alpha_t$, to their measured electroweak-scale values. The resulting partial predictivity for the BSM Yukawa couplings -- especially in $\alpha_{y_1}$ -- is quite strong. Figure~\ref{fig:bsmPartialPredictivity}~shows how the RG flow strongly focuses the BSM Yukawa couplings towards the IR, all the while enforcing that the SM couplings match to their electro-weak scale values. The full range of perturbative EFT values at $\Lambda_\text{Planck}$ is mapped to values below the partial fixed point, i.e., $\alpha_{y_1}(k=\Lambda_\text{ew})\lesssim 0.0165$ and $\alpha_{y_2}(k=\Lambda_\text{ew})\lesssim 0.0083$. These values do not precisely match with the fixed-point values in Eq.~\eqref{eq:predictiveFP} because the SM couplings are matched to their electro-weak scale values, instead.

In general, any RG trajectory for the BSM Yukawa couplings in the gray region of Figure~\ref{fig:bsmPartialPredictivity} is possible. However, typical initial conditions, i.e., those which are not fine-tuned to values very close to zero, are all mapped to values very close to the partial fixed-point value, cf.~thin lines in Figure~\ref{fig:bsmPartialPredictivity}. This is a result of the power-law scaling towards the interacting fixed point in Eq.~\eqref{eq:predictiveFP} (more specifically, towards its partial counterpart). Quantitatively, the RG flow maps initial conditions within the perturbative but `natural' range of coupling values at the Planck scale $\alpha_{y_1}(k=\Lambda_\text{Planck})\,\in\,[10^{-4},0.5]$ to a very narrow window at the electroweak scale, i.e., to $\alpha_{y_1}(k=\Lambda_\text{ew})\,\in\,[1.646\times10^{-2},1.649\times10^{-2}]$. Assuming that the BSM Yukawa couplings should take such `natural', i.e., $\mathcal{O}(1)$, values at the Planck scale, therefore predicts $\alpha_{y_1}(k=\Lambda_\text{ew})\approx1.65\times10^{-2}$. We caution that a correct matching to the SM values of $\alpha_t$ requires the latter to have an `unnatural' Planck scale value $\sim 10^{-5}$, thereby questioning the use of the above naturalness assumption. Similar arguments also apply to $\alpha_{y_2}$, although partial predictivity is less pronounced, cf.~Figure~\ref{fig:bsmPartialPredictivity}.

The above partial predictivity does \emph{not} rely on the existence of a gauge-Yukawa fixed point like the one found in Eq.~\eqref{eq:predictiveFP}. It is merely a consequence of the partial IR fixed-point for the BSM Yukawa couplings, cf.~Eq.~\ref{eq:partialYukawaFP}. We list some explicit examples of BSM matter content to realize the CAF, GY, and CT phase (for both non-Abelian gauge groups) along with predictivity and partial predictivity in Table~\ref{tab:specificModels}. It can be concluded that while only theories in the GY and the CT phase are predictive, partial predictivity persists in all models. In particular, the partial predictivity in the absence of a gauge-Yukawa fixed point can outgrow the partial predictivity in the presence of one. 

\begin{table}
\centering
\renewcommand{\arraystretch}{2}
\begin{tabular}{c|c||c|c|c}
       $(N_F^{R_1},\,d_2^{R_1},\,d_3^{R_1})$ & $(N_F^{R_2},\,d_2^{R_2},\,d_3^{R_2})$
       & predictivity & partial predictivity & non-Abelian phase
       \\\hline\hline
    $(1,\,2,\,1)$ & $(1,\,1,\,3)$  
    & $\mathcal{P}(\Lambda_{NP})=\infty$ & $\widetilde{\mathcal{P}}(\Lambda_{NP})\approx 2\times10^{-4}$
    & CAF
    \\\hline
    $(1,\,3,\,1)$ & $(2,\,1,\,6)$  
    & $\mathcal{P}(\Lambda_{NP})\approx10^{-9}$ & $\widetilde{\mathcal{P}}(\Lambda_{NP})\approx 3\times10^{-4}$
    & GY
    \\\hline
    $(1,\,4,\,1)$ & $(11,\,1,\,3)$  
    & $\mathcal{P}(\Lambda_{NP})\approx10^{-11}$ & $\widetilde{\mathcal{P}}(\Lambda_{NP})\approx 9\times10^{-4}$
    & CT
    \\\hline
    $(1,\,4,\,1)$ & $(1,\,1,\,10)$  
    & $\mathcal{P}(\Lambda_{NP})\approx10^{-12}$ & not viable
    & CT
\end{tabular}
\caption{
\label{tab:specificModels}
Preditictivity $\mathcal{P}(\Lambda_{NP})$ and partial predictivity $\widetilde{\mathcal{P}}(\Lambda_{NP})$ at a new-physics scale $\Lambda_{NP}=10^5\,\text{GeV}$ for some selected BSM models in the three available phases characterized by their BSM matter content in the first two collumns, see main text for further discussion. 
The predictivity for the BSM model content in the last line is already high enough to exclude its validity when matched to SM values of couplings, i.e., by a sub-Planckian Landau pole in $\alpha_3$. (We refrain from listing further examples in mixed phases.)
}
\end{table}

\section{Discussion}
\label{sec:discussion}

We have analyzed the fixed points of gauge-Yukawa theories and, in particular, the SM gauge group in the context of EFTs below the Planck scale. For the SM gauge groups, we have clarified why gauge-Yukawa fixed points with UV-attractive directions cannot occur within the perturbatively controlled regime. However, additional matter fields can result in a perturbative and fully IR-attractive gauge-Yukawa fixed point which realizes \emph{effective asymptotic safety}.
We have introduced a novel quantitative measure for the predictivity of general EFTs and have applied it to gauge-Yukawa BSM extensions.
Concerning concrete BSM phenomenology, this allows us to make the following conclusions:
\begin{itemize}
\item
The results highlight that the presence or absence of the (Abelian) Landau pole is no useful criterion in the search of perturbative interacting fixed points in non-gravitational and hence necessarily effective theories with a Planckian cutoff.
\item
We have identified a fully IR-attractive and (apart from the Abelian gauge coupling) fully interacting fixed point if suitable vector-like fermions without hypercharge, i.e., one $SU(3)$ singlet in the three-dimensional representation of $SU(2)$ and two $SU(2)$ singlets in the six-dimensional representation of $SU(3)$, are added to the SM. This particular theory is predictive along the RG flow towards the IR. We have quantified its predictive power and compared it to other BSM models without interacting fixed points. For all these models, partial predictivity restricts the range of coupling values of the BSM Yukawa couplings in dependence on the ratio between the BSM scale and the cutoff scale.
\item
In general, the predictive power of subplanckian effective asymptotic safety of gauge-Yukawa theories can be estimated by a simple argument: Let $\epsilon\ll 1$ be the perturbative parameter. For simple gauge-Yukawa theories, $\epsilon\lesssim0.1$ has been found in \cite{Litim:2014uca} as the indicated regime of perturbative control. Perturbative fixed points that come about by the balance of loop orders will necessarily result in critical exponents $\theta$ proportional to some power of $\epsilon$, i.e., $\theta\lesssim\epsilon$. Extrapolating the linearized regime around the fixed point, one therefore expects $(\alpha(\Lambda_\text{IR})-\alpha_\ast)/(\alpha(\Lambda_\text{UV})-\alpha_\ast) = \epsilon\,\log(\Lambda_\text{IR}/\Lambda_\text{UV})$ for the associated coupling $\alpha$. For the phenomenologically important case of $\Lambda_\text{NP}/\Lambda_\text{Planck}<\Lambda_\text{ew}/\Lambda_\text{Planck}\sim 10^{17}$, predictivity is thus expected to be limited to shrinking the allowed region of all perturbative coupling values by one or two orders of magnitude. This simple argument also motivates that predictivity can be further increased (i) for non-perturbative fixed points -- as e.g.,~tentatively suggested in a toy model in \cite{Eichhorn:2018vah} -- because $\theta$ need not be small and (ii) for potential fixed points including gravitational fluctuations, see e.g.,~\cite{Shaposhnikov:2009pv,Eichhorn:2017ylw,Eichhorn:2017lry,Eichhorn:2018whv} since $\Lambda_\text{UV}$ can be extended beyond the Planck scale.
\end{itemize}

More generally, the example of gauge-Yukawa theories suggests that the boundaries of all UV-complete and/or IR-complete theories constitute special hypersurfaces in the theory space. In particular, we have made the following observations.
\begin{itemize}
\item
The boundary hypersurfaces separate theories on both sides. Whenever one is confident that such a boundary exists and one knows that experimentally observed values lie either inside or outside, one can exclude that the observed IR physics originates from UV physics on the other side of the boundary.
\item
Moreover, the boundary surfaces can inherit the IR-attractive properties of their delimiting fixed point. In such cases, generic EFTs at the cutoff scale -- both UV complete and not UV complete -- will converge to realize coupling values closer to the boundary surface towards the IR. This is a first step to generalize the local notion of fixed points to global IR-attractors in theory space.
\end{itemize}
These two points highlight that knowledge about such boundary surfaces can be of great value whenever one tries to relate theories at different scales. Of course, having all the information to exactly reconstruct the boundary surface amounts to knowing about all RG flows in its vicinity. One might, therefore, object that with this information one could directly evolve a theory between different scales and obtain its counterpart at other scales. However, this is true only if one knows about \emph{all} the coupling values at a given scale which is typically not the case in the search for new physics. The constraints on BSM Yukawa couplings, that partial predictivity and perturbativity up to the Planck scale entail, provide for an example to emphasize this more general point.

\paragraph*{Acknowledgements:}
The author is grateful to A. Eichhorn for many valuable discussions and comments on the manuscript as well as to C. Nieto for valuable discussion on \cite{Barducci:2018ysr}. Work on this manuscript is supported by the Royal Society International Newton Fellowship NIF\textbackslash R1\textbackslash 191008. 


\appendix
\section*{Appendix}
\renewcommand{\thesection}{\Alph{section}}

\section{General expressions for gauge and Yukawa beta-functions}
\label{app:generalBetas}
For completeness, we collect the literature on expressions for $\overline{\text{MS}}$-scheme 3-loop gauge and 2-loop Yukawa beta-functions of general non-Abelian gauge-Yukawa theories. One can straightforwardly reduce \emph{one} non-Abelian subgroup to an Abelian subgroup. Note that this does \emph{not} directly generalize to multiple Abelian factors because of kinetic mixing, cf.~\cite{delAguila:1988jz,delAguila:1987st}. We neglect all contributions from scalar quartic couplings.
The 1-loop beta-functions for a simple non-Abelian gauge group were first calculated along in \cite{Gross:1973id, Politzer:1973fx} and subsequently generalized to 2-loop and semi-simple groups in \cite{Caswell:1974gg, Tarasov:1976ef, Jones:1981we, Machacek:1983tz}. State-of-the-art 3-loop results have been obtained in \cite{Curtright:1979mg, Pickering:2001aq} for simple and in \cite{Mihaila:2014caa} for semi-simple groups.
For the Yukawa couplings, 1-loop results were first derived in \cite{Cheng:1973nv} and 2-loop results in \cite{Fischler:1982du, Machacek:1983fi, Jack:1983sk}.
Specific results for the Standard Model were derived in \cite{Arason:1991ic, Luo:2002ti, Luo:2002ey} at 2-loop and in \cite{Mihaila:2012fm, Chetyrkin:2012rz, Bednyakov:2012en, Bednyakov:2013eba} at 3-loop order. Neglecting contributions from other couplings the simple gauge beta-functions have been calculated up to 5-loop order \cite{Baikov:2016tgj, Herzog:2017ohr, Luthe:2017ttg}. Finally, SARAH \cite{Staub:2012pb, Staub:2013tta} and PyR@TE \cite{Lyonnet:2013dna, Lyonnet:2016xiz} provide computer algebra tools for beta-functions at two-loop level, see also \cite{Schienbein:2018fsw}.

For the present purpose, it is sufficient to work with the 3-loop gauge beta functions (including general Yukawa-coupling contributions but neglecting quartic couplings) and with the 2-loop Yukawa beta functions (again neglecting quartic couplings).
Focusing on a simple gauge group with and one fermionic representation $R_F$, the Lagrangian reads
\begin{align}
\mathcal{L} = &-\frac{1}{4} F_A^{\mu\nu} F^A_{\mu \nu}  + \mathcal{L}_{(R_F)} \quad\text{with}
\notag\\
\mathcal{L}_{(R_F)} =&+ \frac{1}{2} D^\mu \phi_a D_\mu \phi_a + i \psi_j^\dagger \sigma^\mu D_\mu \psi_j
- \frac{1}{2} \left(Y^a_{jk} \psi_j \zeta \psi_k \phi_a + Y^{a*}_{jk} \psi_j^\dagger \zeta \psi_k^\dagger \phi_a\right)\;.
\label{eq:generalLagrangian}
\end{align}
Here, we have already assumed that each fermionic representation $R_F$ is accompanied by a suitable set of scalars to facilitate Yukawa couplings $Y^a_{ij}$. 
Here, $F^A_{\mu \nu} = \partial_\mu A_\nu - \partial_\nu A_\mu + g\,f^{ABC} A_\mu^b A_\nu^c$ is the standard field-strength tensor with gauge group structure constants $f^{ABC}$ and $\zeta = i\sigma_2$ with $\sigma_2$ the 2nd Pauli matrix. Fermions and scalars are minimally coupled via covariant derivatives corresponding with generators $t_{ij}^{A}$ and $\theta_{ab}^{A}$, respectively, i.e.,
\begin{align}
    D_\mu\phi_a &= \partial_\mu\phi_a - i\,g\,\theta_{ab}^A\,A_\mu^{A}\,\phi_b\;,
    \\
    D_\mu\psi_i &= \partial_\mu\psi_i - i\,g\,t_{ij}^A\,A_\mu^{A}\,\psi_j\;.    
\end{align}
Finally, $Y^a_{ij}$ denote complex Yukawa coupling matrices.
A generalization to multiple gauge groups and representations is straightforward. We have omitted quartic couplings and mass terms since we neglect them for this paper.

Most of the contributions to the beta-functions can be written in terms of the respective quadratic Casimirs $C_2$ and Dynkin indices $S_2$. For the adjoint gauge fields, fermions, and scalars they read
\begin{align}
    C_2^\text{adj}\delta^{AB} &= f^{ACD}f^{BCD}\;,
    &C_2^{R_F} &= \sum_{A=1}^d t^A t^A\;,
    &C_2^{R_S} &= \sum_{A=1}^d \theta^A \theta^A\;, 
    \\
    &&S_2^{R_F} \delta^{AB}&=\Tr[t^A t^B]\;,
    &S_2^{R_S} \delta^{AB}&=\Tr[\theta^A \theta^B]\;.
\end{align}
We denote the associated beta-functions by
\begin{align}
    \beta_{g} &= \left[\frac{\beta_g^\text{(1-loop)}}{(4\pi)^2} + \frac{\beta_g^\text{(2-loop)}}{(4\pi)^4} + \frac{\beta_g^\text{(3-loop)}}{(4\pi)^6}\right]\;,
    \\
    \beta_{Y^a} &= \left[\frac{\beta_a^\text{(1-loop)}}{(4\pi)^2} + \frac{\beta_a^\text{(2-loop)}}{(4\pi)^4}\right]\;.
\end{align} 
The explicit expressions can be found in \cite{Pickering:2001aq} and \cite{Schienbein:2018fsw}, respectively. 
Note, that we omit contributions from scalar quartic couplings. 
The replacement rules to generalize to semi-simple gauge groups at 2-loop level can be found in \cite{Luo:2002ti}.
All the required beta-functions can thereby be derived from this general set of expressions.

\section{Reduction to (B)SM beta-functions}
\label{app:betasForBSM}
To obtain the beta-functions for the BSM extensions covered in Section~\ref{sec:newMatter}, we specialize to the SM Lagrangian supplemented by $\mathcal{N}$ different types of vector-like BSM fermions $\Psi^{(I)}$ in the representation $R_I$, where $I=1,\,\dots,\,\mathcal{N}$ labels the different types of BSM representations, each of which can come with a multiplicity $N_{R_I}$. For each such type of vectorlike BSM fermions, we also include an $N_{R_I}\times N_{R_I}$ matrix of uncharged complex scalars $\phi_{(I)}$ which allows for Yukawa couplings. The corresponding Lagrangian reads
\begin{align}
\label{eq:BSMLagrangian}
    \mathcal{L} = \mathcal{L}_\text{SM} 
    + \sum_{I=1}^{\mathcal{N}}
    \Bigg[
        \text{Tr}\left(\overline{\psi}^{(I)}i\slashed{D}\psi^{(I)}\right)
        + \text{Tr}\left(\partial_\mu\phi_{(I)}^\dagger\partial^\mu\phi_{(I)}\right)
        + y_{I}\text{Tr}\left(\overline{\psi}^{(I)}_L\phi_{(I)}\psi^{(I)}_R + \overline{\psi}^{(I)}_R\phi_{(I)}^\dagger\psi^{(I)}_L\right)
    \Bigg]\;,
\end{align}
where we have decomposed the BSM fermions into two-component left- and right-handed parts, i.e., $\psi^{(I)}_{R/L} = \frac{1}{2}(1\pm\gamma_5)\psi^{(I)}$. We neglect any quartic couplings. Therefore, the different types of BSM fermions couple to each other as well as to the SM only via their minimal gauge interactions.
The matrix-Yukawa couplings in Eq.~\eqref{eq:generalLagrangian} are diagonal and therefor the different traces of Yukawa-matrices in the beta-functions, cf.~\cite{Pickering:2001aq,Schienbein:2018fsw}, simplify.
Introducing, for the gauge couplings $\alpha_{i} = \frac{g_i^2}{(4\pi)^2}$ for $i=1,\,2,\,3$, for the top-Yukawa coupling $\alpha_{t} = \frac{y_{t}^2}{(4\pi)^2}$, and for the BSM Yukawa couplings $\alpha_{I} = \frac{y_{(I)}^2}{(4\pi)^2}$, the general NLO and NNLO contributions to the beta-functions reduce to the follwing form
\begin{align}
    \beta_{\alpha_i}^\text{(NLO)} &= \alpha_i^2\Bigg[
        B_{i} + \sum_{j=1}^3 C_{ij}\alpha_j - \sum_{I=0}^{\mathcal{N}} D_{iI}\alpha_I
    \Bigg]\;,
    \\
    \beta_{\alpha_I}^\text{(NLO)} &= \alpha_I\Bigg[
        E_I \alpha_I - \sum_{j=1}^{3}F_{Ij}\alpha_j
    \Bigg]\;,
    \\
    \beta_{\alpha_i}^\text{(NNLO)} &= \alpha_i^2\Bigg[
        \sum_{j=1}^3\sum_{k=j}^3 M_{ijk}\alpha_j\alpha_k 
        - \sum_{j=1}^3\sum_{I=0}^{\mathcal{N}} K_{ijI}\alpha_j\alpha_I
        + \sum_{I=0}^{\mathcal{N}} \overline{K}_{iI}\alpha_I^2
    \Bigg]\;,
    \\
    \beta_{\alpha_I}^\text{(NNLO)} &= \alpha_I\Bigg[
        \sum_{j=1}^{3}V_{Ij}\alpha_j\alpha_I + \overline{V}_I\alpha_I^2
        + \sum_{i=1}^3\sum_{j=i}^3W_{Ijk}\alpha_j\alpha_k        
    \Bigg]\;,
\end{align}
where the notation includes the SM top-Yukawa coupling as $\alpha_t = \alpha_{I=0}$. The group theoretic coefficients $B_{i}$, $C_{ij}$, $D_{iI}$, $E_I$, $F_{Ij}$ at NLO level and $M_{ijk}$, $K_{ijI}$, $\overline{K}_{iI}$, $V_{Ij}$, $\overline{V}_I$, $W_{Ijk}$ at NNLO level are well-known for the SM \cite{Arason:1991ic, Luo:2002ti, Luo:2002ey, Mihaila:2012fm, Chetyrkin:2012rz, Bednyakov:2012en, Bednyakov:2013eba}. The BSM contributions can straightforwardly be generalized from the expressions for a single type of BSM representation, cf.~\cite{Barducci:2018ysr}, by suitable summation. For completeness, we list them below, where the Casimirs $C_{N_C}^{R_I}$, Dynkin indices $S_{N_C}^{R_I}$, and dimensions $d_{N_C}^{R_I}$ for the BSM representations $R_I$ of the non-Abelian SM gauge groups $SU(N_C)$ with $N_C=2$ and $N_C=3$ are given by
\begin{align}
    C_{2}^{R_I}&=\ell(\ell+1)\;,
    &d_{2}^{R_I}&=2\ell+1\;,
    &S_{2}^{R_I}&=\frac{C_{2}^{R_I}d_{2}^{R_I}}{3}\;,
    \label{eq:groupTheoryData_SU2}
    \\
    C_{3}^{R_I}&=\frac{1}{2}(p+1)(q+1)(p+q+2)\;,
    &d_{3}^{R_I}&=p+q+\frac{1}{3}\left(p^2+q^2+p\,q\right)\;,
    &S_{3}^{R_I}&=\frac{C_{3}^{R_I}d_{3}^{R_I}}{8}\;,
    \label{eq:groupTheoryData_SU3}
\end{align}
where $l=1/2,\,1,\,3/2,\,\dots$ and $p,q = 1,\,2,\,3,\,\dots$ denote the highest weights for $SU(2)$ and $SU(3)$, respectively.

The group-theoretic NLO, i.e., 1-loop and 2-loop, coefficients in the gauge beta functions of the SM including BSM representations read
\begin{align*}
    B_{1}&=\frac{41}{3}+\frac{8}{3}\sum_{I=1}^\mathcal{N}N_{R_I}Y_{R_I}^{2}d^{R_I}_{2}d^{R_I}_{3}
    &B_{2}&=-\frac{19}{3}+\frac{8}{3}\sum_{I=1}^\mathcal{N}N_{R_I}S_2^{R_I}d^{R_I}_3
    \\
    B_{3}&=-14+\frac{8}{3}\sum_{I=1}^\mathcal{N}N_{R_I}S^{R_I}_{3}d^{R_I}_{2}
    &C_{11}&=\frac{199}{9}+8\sum_{I=1}^\mathcal{N}Y_{R_I}^{4}N_{R_I}d^{R_I}_{2}d^{R_I}_{3}
    \\
    C_{12}&=9+8\sum_{I=1}^\mathcal{N}Y_{R_I}^{2}N_{R_I}C^{R_I}_{2}d^{R_I}_{2}d^{R_I}_{3}
    &C_{13}&=\frac{88}{3}+8\sum_{I=1}^\mathcal{N}N_{R_I}Y_{R_I}^{2}C^{R_I}_{3}d^{R_I}_{2}d^{R_I}_{3}
    \\
    C_{21}&=3+8\sum_{I=1}^\mathcal{N}N_{R_I}Y_{R_I}^{2}S^{R_I}_{2}d^{R_I}_{3}
    &C_{22}&=\frac{35}{3}+4\sum_{I=1}^\mathcal{N}N_{R_I}S^{R_I}_{2}d^{R_I}_{3}\left(2\, C^{R_I}_{2}+\frac{20}{3}\right)
    \\
    C_{23}&=24+8\sum_{I=1}^\mathcal{N}N_{R_I}S^{R_I}_{2}C^{R_I}_{3}d^{R_I}_{3}
    &C_{31}&=\frac{11}{3}+8\sum_{I=1}^\mathcal{N}N_{R_I}Y_{R_I}^{2}S^{R_I}_{3}d^{R_I}_{2}
    \\
    C_{32}&=9+8\sum_{I=1}^\mathcal{N}N_{R_I}S^{R_I}_{3}C^{R_I}_{2}d^{R_I}_{2}
    &C_{33}&=-52+4\sum_{I=1}^\mathcal{N}N_{R_I}S^{R_I}_{3}d^{R_I}_{2}(2\,C^{R_I}_{3}+10)
    \\
    D_{1I}&=4\sum_{I=1}^\mathcal{N}N_{R_I}^{2}Y_{R_I}^{2}d^{R_I}_{2}d^{R_I}_{3}
    &D_{2I}&=\frac{1}{3}4\sum_{I=1}^\mathcal{N}N_{R_I}^{2}C^{R_I}_{2}d^{R_I}_{2}d^{R_I}_{3}
    \\
    D_{3I}&=\frac{1}{8}4\sum_{I=1}^\mathcal{N}N_{R_I}^{2}C^{R_I}_{3}d^{R_I}_{2}d^{R_I}_{3}
\end{align*}
The group-theoretic NLO, i.e., 1-loop, coefficients in the Yukawa beta functions of the SM including BSM representations read
\begin{align*}
    E_t &= 9
    &E_I &= 2(N_{R_I}+d^{R_I}_{2}d^{R_I}_{3}) 
    \\
    F_{t1}&=\frac{17}{6}
    &F_{t2}&=\frac{9}{2}
    &F_{t3}&=16
    \\
    F_{I1}&= 12Y_{R_I}^2
    &F_{I2}&= 12C^{R_I}_{2}
    &F_{I3}&= 12C^{R_I}_{3}
\end{align*}
The group-theoretic NNLO, i.e., 3-loop, coefficients in the gauge beta functions of the SM including BSM representations read
\begin{align*}
    M_{111}&= \frac{388613}{2592}+\sum_{I=1}^\mathcal{N}\Bigg[
    \frac{4405}{162}N_{R_I}Y_{R_I}^{2}d^{R_I}_{2}d^{R_I}_{3}+\frac{463}{9}N_{R_I}Y_{R_I}^{4}d^{R_I}_{2}d^{R_I}_{3}
+4N_{R_I}Y_{R_I}^{6}d^{R_I}_{2}d^{R_I}_{3}+\frac{88}{9}N_{R_I}^{2}Y_{R_I}^{6}\left(d^{R_I}_{2}d^{R_I}_{3}\right)^{2}\Bigg]
    \\
    M_{122}&=\frac{1315}{32}+\sum_{I=1}^\mathcal{N}\Bigg[\frac{245}{9}C^{R_I}_{2}N_{R_I}Y_{R_I}^{2}d^{R_I}_{2}d^{R_I}_{3}-4\, \left(C^{R_I}_{2}\right)^{2}N_{R_I}Y_{R_I}^{2}d^{R_I}_{2}d^{R_I}_{3}+\frac{23}{2}N_{R_I}S^{R_I}_{2}d^{R_I}_{3}\Bigg]
    \\
    M_{133}&=198+\sum_{I=1}^\mathcal{N}\Bigg[\frac{178}{3}C^{R_I}_{3}N_{R_I}Y_{R_I}^{2}d^{R_I}_{2}d^{R_I}_{3}-4\, \left(C^{R_I}_{3}\right)^{2}N_{R_I}Y_{R_I}^{2}d^{R_I}_{2}d^{R_I}_{3}-\frac{968}{27}N_{R_I}S^{R_I}_{3}d^{R_I}_{2}\Bigg]
    \\
    M_{112}&=\frac{205}{48}-\sum_{I=1}^\mathcal{N}8\, C^{R_I}_{2}N_{R_I}Y_{R_I}^{4}d^{R_I}_{2}d^{R_I}_{3}
    \\
    M_{113}&=\frac{274}{27}+\sum_{I=1}^\mathcal{N}8\, C^{R_I}_{3}N_{R_I}Y_{R_I}^{4}d^{R_I}_{2}d^{R_I}_{3}
    \\
    M_{123}&=2+\sum_{I=1}^\mathcal{N}8\, C^{R_I}_{2}C^{R_I}_{3}N_{R_I}Y_{R_I}^{2}d^{R_I}_{2}d^{R_I}_{3}
    \\
    M_{211}&=\frac{5597}{288}+\sum_{I=1}^\mathcal{N}\Bigg[\frac{23}{6}N_{R_I}Y_{R_I}^{2}d^{R_I}_{2}d^{R_I}_{3}+\frac{463}{9}Y_{R_I}^{2}N_{R_I}S^{R_I}_{2}d^{R_I}_{3}+4N_{R_I}Y_{R_I}^{4}S^{R_I}_{2}d^{R_I}_{3}\Bigg]
    \\
    M_{222}&=\frac{324953}{864}+\sum_{I=1}^\mathcal{N}\Bigg[\frac{13411}{54}N_{R_I}S^{R_I}_{2}d^{R_I}_{3}+\frac{533}{9}N_{R_I}C^{R_I}_{2}S^{R_I}_{2}d^{R_I}_{3}-4N_{R_I}\left(C^{R_I}_{2}\right)^{2}S^{R_I}_{2}d^{R_I}_{3}\Bigg]
    \\
    M_{233}&=162+\sum_{I=1}^\mathcal{N}\Bigg[\frac{178}{3}C^{R_I}_{3}N_{R_I}S^{R_I}_{2}d^{R_I}_{3}-4\, \left(C^{R_I}_{3}\right)^{2}N_{R_I}S^{R_I}_{2}d^{R_I}_{3}-\frac{88}{3}N_{R_I}S^{R_I}_{3}d^{R_I}_{2}\Bigg]
    \\
    M_{212}&=\frac{291}{16}+\sum_{I=1}^\mathcal{N}\Bigg[32\, Y_{R_I}^{2}N_{R_I}S^{R_I}_{2}d^{R_I}_{3}-8\, Y_{R_I}^{2}C^{R_I}_{2}N_{R_I}S^{R_I}_{2}d^{R_I}_{3}\Bigg]
    \\
    M_{213}&=\frac{2}{3}+\sum_{I=1}^\mathcal{N}8\, Y_{R_I}^{2}C^{R_I}_{3}N_{R_I}S^{R_I}_{2}d^{R_I}_{3}
    \\
    M_{223}&=78+\sum_{I=1}^\mathcal{N}\Bigg[32\, C^{R_I}_{3}N_{R_I}S^{R_I}_{2}d^{R_I}_{3}-8\, C^{R_I}_{2}C^{R_I}_{3}N_{R_I}S^{R_I}_{2}d^{R_I}_{3}\Bigg]
    \\
    M_{211}&=\frac{5597}{288}+\sum_{I=1}^\mathcal{N}\Bigg[\frac{23}{6}N_{R_I}Y_{R_I}^{2}d^{R_I}_{2}d^{R_I}_{3}+\frac{463}{9}Y_{R_I}^{2}N_{R_I}S^{R_I}_{2}d^{R_I}_{3}+4N_{R_I}Y_{R_I}^{4}S^{R_I}_{2}d^{R_I}_{3}\Bigg]
    \\
    M_{311}&=\frac{2615}{108}+\sum_{I=1}^\mathcal{N}\Bigg[\frac{121}{27}N_{R_I}Y_{R_I}^{2}d^{R_I}_{2}d^{R_I}_{3}+\frac{463}{9}Y_{R_I}^{2}N_{R_I}S^{R_I}_{3}d^{R_I}_{2}+4N_{R_I}Y_{R_I}^{4}S^{R_I}_{3}d^{R_I}_{2}\Bigg]
    \\
    M_{322}&=\frac{109}{4}+\sum_{I=1}^\mathcal{N}\Bigg[-11N_{R_I}S^{R_I}_{2}d^{R_I}_{3}+\frac{245}{9}C^{R_I}_{2}N_{R_I}S^{R_I}_{3}d^{R_I}_{2}-4\, \left(C^{R_I}_{2}\right)^{2}N_{R_I}S^{R_I}_{3}d^{R_I}_{2}\Bigg]
    \\
    M_{333}&=65+\sum_{I=1}^\mathcal{N}\Bigg[\frac{6242}{9}N_{R_I}S^{R_I}_{3}d^{R_I}_{2}+\frac{322}{3}N_{R_I}C^{R_I}_{3}S^{R_I}_{3}d^{R_I}_{2}-4N_{R_I}\left(C^{R_I}_{3}\right)^{2}S^{R_I}_{3}d^{R_I}_{2}\Bigg]
    \\
    M_{312}&=\frac{1}{4}+\sum_{I=1}^\mathcal{N}8\, Y_{R_I}^{2}C^{R_I}_{2}N_{R_I}S^{R_I}_{3}d^{R_I}_{2}
    \\
    M_{313}&=\frac{154}{9}+\sum_{I=1}^\mathcal{N}\Bigg[48\, Y_{R_I}^{2}N_{R_I}S^{R_I}_{3}d^{R_I}_{2}-8\, Y_{R_I}^{2}C^{R_I}_{3}N_{R_I}S^{R_I}_{3}d^{R_I}_{2}\Bigg]
    \\
    M_{323}&=42+\sum_{I=1}^\mathcal{N}\Bigg[48\, C^{R_I}_{2}N_{R_I}S^{R_I}_{3}d^{R_I}_{2}-8\, C^{R_I}_{2}C^{R_I}_{3}N_{R_I}S^{R_I}_{3}d^{R_I}_{2}\Bigg]
\end{align*}
as well as
\begin{align*}
    K_{11I}&=6\, Y_{R_I}^{4}N_{R_I}^{2}d^{R_I}_{2}d^{R_I}_{3}
    &K_{12I}&=6\, Y_{R_I}^{2}C^{R_I}_{2}N_{R_I}^{2}d^{R_I}_{2}d^{R_I}_{3}
    \\
    K_{13I}&=6\, Y_{R_I}^{2}C^{R_I}_{3}N_{R_I}^{2}d^{R_I}_{2}d^{R_I}_{3}
    &K_{21I}&=2\, Y_{R_I}^{2}C^{R_I}_{2}N_{R_I}^{2}d^{R_I}_{2}d^{R_I}_{3}
    \\
    K_{22I}&=16\, C^{R_I}_{2}N_{R_I}^{2}d^{R_I}_{2}d^{R_I}_{3}+2\, \left(C^{R_I}_{2}\right)^{2}N_{R_I}^{2}d^{R_I}_{2}d^{R_I}_{3}
    &K_{23I}&=2\, C^{R_I}_{2}C^{R_I}_{3}N_{R_I}^{2}d^{R_I}_{2}d^{R_I}_{3}
    \\
    K_{31I}&=\frac{3}{4}Y_{R_I}^{2}C^{R_I}_{3}N_{R_I}^{2}d^{R_I}_{2}d^{R_I}_{3}
    &K_{32I}&=\frac{3}{4}C^{R_I}_{2}C^{R_I}_{3}N_{R_I}^{2}d^{R_I}_{2}d^{R_I}_{3}
    \\
    K_{33I}&=9\, C^{R_I}_{3}N_{R_I}^{2}d^{R_I}_{2}d^{R_I}_{3}+\frac{3}{4}\left(C^{R_I}_{3}\right)^{2}N_{R_I}^{2}d^{R_I}_{2}d^{R_I}_{3}
    &\overline{K}_{1I}&=6\, Y_{R_I}^{2}N_{R_I}^{3}d^{R_I}_{2}d^{R_I}_{3}+7Y_{R_I}^{2}N_{R_I}^{2}\left(d^{R_I}_{2}d^{R_I}_{3}\right)^{2}
    \\
    \overline{K}_{2I}&=2\, C^{R_I}_{2}N_{R_I}^{3}d^{R_I}_{2}d^{R_I}_{3}+\frac{7}{3}C^{R_I}_{2}N_{R_I}^{2}\left(d^{R_I}_{2}d^{R_I}_{3}\right)^{2}
    &\overline{K}_{3I}&=\frac{3}{4}C^{R_I}_{3}N_{R_I}^{3}d^{R_I}_{2}d^{R_I}_{3}+\frac{7}{8}C^{R_I}_{3}N_{R_I}^{2}\left(d^{R_I}_{2}d^{R_I}_{3}\right)^{2}
    \\
    K_{11t}&=\frac{2827}{144}
    \quad\quad\quad\quad\quad
    K_{12t}=\frac{785}{16}
    &K_{13t}&=\frac{58}{3}
    \quad\quad\quad\quad\quad
    K_{21t}=\frac{593}{48}
    \\
    K_{22t}&=\frac{729}{16}
    \quad\quad\quad\quad\quad
    K_{23t}=14
    &K_{31t}&=\frac{101}{12}
    \quad\quad\quad\quad\quad
    K_{32t}=\frac{93}{4}
    \\
    K_{33t}&=80
    \quad\quad\quad\quad\quad
    \overline{K}_{1t}=\frac{315}{8}
    &\overline{K}_{2t}&=\frac{147}{8}
    \quad\quad\quad\quad\quad
    \overline{K}_{3t}=30
\end{align*}
The group-theoretic NNLO, i.e., 2-loop, coefficients in the Yukawa beta functions of the SM including BSM representations read
\begin{align*}
    V_{I1}&=2\, (8\, N_{R_I}+5\, d^{R_I}_{2}d^{R_I}_{3})Y_{R_I}^{2}
    &V_{I2}&=2\, (8\, N_{R_I}+5\, d^{R_I}_{2}d^{R_I}_{3})C^{R_I}_{2}
    \\
    V_{I3}&=2\, (8\, N_{R_I}+5\, d^{R_I}_{2}d^{R_I}_{3})C^{R_I}_{3}
    &\overline{V}_{I}&=4 - \frac{1}{2}N_{R_I}^{2}+3\, N_{R_I}d^{R_I}_{2}d^{R_I}_{3}
    \\
    W_{I11}&=\left(\frac{211}{3}-6Y_{R_I}^{2}+\frac{40}{3}Y_{R_I}^{2}N_{R_I}d^{R_I}_{2}d^{R_I}_{3}\right)Y_{R_I}^{2}
    &W_{I22}&=\left(-\frac{257}{3}-6C^{R_I}_{2}+\frac{40}{3}N_{R_I}S^{R_I}_{2}d^{R_I}_{3}\right)C^{R_I}_{2}
    \\
    W_{I33}&=\left(-154-6C^{R_I}_{3}+\frac{40}{3}N_{R_I}S^{R_I}_{3}d^{R_I}_{2}\right)C^{R_I}_{3}
    &W_{I12}&=-12\, Y_{R_I}^{2}C^{R_I}_{2}
    \\
    W_{I13}&=-12\, Y_{R_I}^{2}C^{R_I}_{3}
    &W_{I23}&=-12\, C^{R_I}_{2}C^{R_I}_{3}
    \\
    V_{t1}&=\frac{131}{8}
    &V_{t2}&=\frac{225}{8}
    \\
    V_{t3}&=72
    &\overline{V}_{t}&=-24
    \\
    W_{t11}&=\frac{1187}{108} + \frac{58}{27}Y_{R_I}^{2}N_{R_I}d^{R_I}_{2}d^{R_I}_{3}
    &W_{t22}&=-\frac{23}{2} + 2S^{R_I}_{2}N_{R_I}d^{R_I}_{3}
    \\
    W_{t33}&=-216 + \frac{160}{9}S^{R_I}_{3}N_{R_I}d^{R_I}_{2}
    &W_{t12}&=\frac{3}{2}
    \\
    W_{t13}&=\frac{38}{9}
    &W_{t23}&=18
\end{align*}

\section{Coefficients for simple gauge groups}
\label{app:U1coeffs}
General expressions for the beta functions of the simple gauge groups of the SM with a single type of BSM matter, i.e., $\mathcal{N}=1$ in Eq.~\ref{eq:BSMLagrangian}, can be extracted from the SM expressions in App.~\ref{app:betasForBSM}. We denote the beta function coefficients by
\begin{align}
\label{eq:simpleGaugeBeta}
    \beta_{\alpha_i}^\text{(NLO)} &= \alpha_g^2\Big[
        B + C\alpha_g - D\alpha_y
    \Big]\;,
    &\beta_{\alpha_i}^\text{(NNLO)} &= \alpha_g^2\Big[
        M\,\alpha_g^2 
        - K\,\alpha_g\alpha_y
        + \overline{K}\,\alpha_y^2
    \Big]\;,
    \\
\label{eq:simpleYukawaBeta}
    \beta_{\alpha_y}^\text{(NLO)} &= \alpha_y\Big[
        E\,\alpha_y - F\,\alpha_y
    \Big]\;,
    &\beta_{\alpha_y}^\text{(NNLO)} &= \alpha_y\Big[
        V\,\alpha_g\alpha_y + \overline{V}\,\alpha_y^2
        + W\,\alpha_g^2        
    \Big]\;,
\end{align}
The coefficients can be obtained by subtracting the SM part, e.g.,~those of U(1) by
\begin{align}
    B &\equiv B_1 - B_1|_{N_{R_I}=0}
    =\frac{8}{3}\,N_{F}Y^{2}
    \;,\\
    C &\equiv C_{11} - C_{11}|_{N_{R_I}=0}
    =8\,Y^{4}N_F
    \;,\\
    D &\equiv D_{1I} - D_{1I}|_{N_{R_I}=0}
    = 4N_F^{2}Y^{2}
    \;,\\
    E &\equiv E_I - E_I|_{N_{R_I}=0}
    = 2(N_F+1)
    \;,\\
    F &\equiv F_{I1} - F_{I1}|_{N_{R_I}=0}
    = 12Y^2
    \;,\\
    M &\equiv M_{111} - B_{111}|_{N_{R_I}=0}
    =     \frac{4405}{162}N_F Y^{2}
    +\frac{463}{9}N_F Y^{4}
    +4N_F Y^{6}
    +\frac{88}{9}N_F^{2}Y^{6}
    \;,\\
    K &\equiv K_{11I} - K_{11I}|_{N_{R_I}=0}
    = 6\, Y^{4}N_F^{2}
    \;,\\
    \overline{K} &\equiv \overline{K}_{1I} - \overline{K}_{1I}|_{N_{R_I}=0}
    = 6\, Y^{2}N_F^{3}+7Y^{2}N_F^{2}
    \;,
\end{align}
where we have reduced to singlets under the other gauge groups, i.e., $d^{R_I}_{2}=1=d^{R_I}_{3}$, and replaced the notation $N_{R_I}\rightarrow N_F$ as well as $Y_{R_I}\rightarrow Y$.

\bibliographystyle{frontiersinHLTHFPHY}
\bibliography{nfAS_notes}
    
\end{document}